\documentstyle[11pt,epsfig]{article}

\DeclareMathAlphabet   {\mathsc}{OT1}{cmr}{m}{sc}

\def\[{\left [}
\def\]{\right ]}
\def\({\left (}
\def\){\right )}
\newcommand{\lang}{\left\langle}
\newcommand{\rang}{\right\rangle}
\newcommand{\lbr}{\left\{}
\newcommand{\rbr}{\right\}}
\newcommand{\oline}[1]{\overline{#1}}

\newcommand{\wtd}[1]{\widetilde{#1}}

\newcommand{\wh}[1]{\widehat{#1}}
\newcommand{\h}[1]{\hat{#1}}

\newcommand{\GeV}      {~\mathrm{GeV}}
\newcommand{\TeV}      {~\mathrm{TeV}}

\newcommand{\SM}       {\mathsc{sm}}

\newcommand{\EM}       {\mathsc{em}}

\newcommand{\UV}       {\mathsc{uv}}
\newcommand{\GS}       {\mathsc{gs}}
\newcommand{\PL}       {\mathsc{pl}}
\newcommand{\PV}       {\mathsc{pv}}
\newcommand{\GUT}      {\mathsc{gut}}
\newcommand{\STR}      {\mathsc{str}}
\newcommand{\SUSY}     {\mathsc{susy}}

\newcommand{\supal}[1]{{#1}^{\alpha}}

\newcommand{\hc}       {\mathrm{\; h.c. \;}}

\newcommand{\order}{{\cal O}}
\newcommand{\re}{{\rm Re}}

\newcommand{\gappeq}{\mathrel{\rlap {\raise.5ex\hbox{$>$}}
{\lower.5ex\hbox{$\sim$}}}}
\newcommand{\lappeq}{\mathrel{\rlap{\raise.5ex\hbox{$<$}}
{\lower.5ex\hbox{$\sim$}}}}

\def\Eisen{G_{2}\(t,\bar{t}\)}

\def\Eisen{G_{2}\(t,\bar{t}\)}

\begin{document}

\begin{flushright}
LBNL--53419 \\ UCB-PTH-03/17 \\ LPT Orsay--03/54 \\ MCTP - 03/36 \\
hep-ph/0308047
\end{flushright}
\begin{center}
{\Large \bf Phenomenological Aspects of Heterotic Orbifold Models
\\ at One Loop}
\vspace{0.8cm}

{\sc Pierre Bin\'etruy$^a$, Andreas Birkedal-Hansen$^b$, Yann
Mambrini$^a$ and \\ Brent D. Nelson$^c$}

\vspace{0.5cm} {\it $^a$ Laboratoire de Physique Th\'eorique\footnote{Unit\'e
Mixte de Recherche UMR $n^{o}$ 8627 - CNRS}, \\ Universit\'e Paris-Sud, F-91405 Orsay, France}

\vspace{0.5cm} {\it$^b$ Department of Physics and Theoretical
Physics Group, Lawrence Berkeley Laboratory, \\ University of
California, Berkeley, CA 94720, USA }

\vspace{0.5cm} {\it$^c$ Michigan Centre for Theoretical Physics,
Randall Laboratory,\\

University of Michigan, Ann Arbor MI 48109 }

\end{center}
\vspace{1cm}

\begin{abstract}

We provide a detailed study of the phenomenology of orbifold
compactifications of the heterotic string within the context of
supergravity effective theories. Our investigation focuses on
those models where the soft Lagrangian is dominated by loop
contributions to the various soft supersymmetry breaking
parameters. Such models typically predict non-universal soft
masses and are thus significantly different from minimal
supergravity and other universal models. We consider the pattern
of masses that are governed by these soft terms and investigate
the implications of certain indirect constraints on supersymmetric
models, such as flavor-changing neutral currents, the anomalous
magnetic moment of the muon and the density of thermal relic
neutralinos. These string-motivated models show novel behavior
that interpolates between the phenomenology of unified
supergravity models and models dominated by the superconformal
anomaly.

\end{abstract}
\newpage

\section*{Introduction}
\label{sec:intro}

The recent interest in models where supersymmetry breaking is
transmitted from a hidden sector to our observable world via the
superconformal anomaly~\cite{GiLuMuRa98,RaSu99,PoRa99} has served
as a reminder that there are occasions where working at one loop
order is more than a theoretical luxury, but an absolute necessity
if one is to understand even the gross features of some models.
Recently the complete one loop supergravity correction to soft
supersymmetry breaking terms has been
obtained~\cite{GaNeWu99,GaNe00b}, a subset of which are the
aforementioned ``anomaly-mediated'' terms. The broad features of
these corrections, in particular for gaugino masses, was studied
in a heterotic string context in~\cite{BiGaNe01}. In the current
work we wish to study in greater detail the low energy
phenomenology of models based on orbifold compactifications of the
weakly-coupled heterotic string in which these loop corrections
are significant. To that extent we will choose two
``laboratories'' in which tree level soft supersymmetry breaking
terms are either absent or greatly suppressed.

Our first example will involve an observable sector in which the
matter fields are assumed to be from the untwisted sector of the
orbifold compactification. In the event that supersymmetry
breaking is transmitted by the compactification moduli $T^\alpha$,
whose vacuum expectation values determine the size of the compact
manifold, then this model will exhibit a no-scale pattern of soft
terms with vanishing soft supersymmetry breaking terms at tree
level. Clearly the phenomenology of such models will depend
critically on the form of the loop induced soft terms. Indeed, the
``sequestered sector'' models considered in~\cite{RaSu99} were of
just such a form.

In our second example we will switch our focus to cases in which
it is the dilaton field $S$, whose vacuum expectation value
determines the magnitude of the (unified) coupling constant
$g_{\STR}$ at the string scale, that participates exclusively in
supersymmetry breaking. We will work in the context of models in
which string nonperturbative corrections to the K\"ahler potential
act to stabilize the dilaton in the presence of gaugino
condensation~\cite{BiGaWu96,BiGaWu97a}. Such theories predict a
suppression of dilaton contributions to soft terms relative to
those of the supergravity auxiliary field -- thereby suppressing
tree level contributions to supersymmetry breaking relative to
certain loop contributions.

After describing our framework and introducing both the tree level
and one loop soft supersymmetry breaking terms in
Section~\ref{sec:general}, we introduce the parameter space for
our two broad classes of models in Section~\ref{sec:space}. The
subsequent section describes the analysis tools we employ and the
types of low-energy phenomena we will study before describing the
implications of low-energy experimental constraints on each of our
models in Section~\ref{sec:phenom}.

In general we find that these examples have a phenomenology that
combines the features of minimal supergravity (mSUGRA) models with
those of anomaly-mediated models. Parameters related to the
orbifold compactification and the stabilization of string moduli
interpolate between these regimes. Current observations and
particle mass limits from collider experiments already shed light
on the nature of moduli stabilization. Future discoveries of
superpartners will likely provide strong evidence in favor of one
stabilization mechanism or another in the context of
weakly-coupled heterotic string models.

\section{The structure of heterotic orbifold models at one loop}
\label{sec:general}

We work in a framework in which the chiral superfields $Z^M$
can be divided into two classes: observable
sector superfields, noted $Z^i$, charged under the observable sector gauge
symmetries; and hidden sector
superfields, noted $Z^n$. Since our interest here is primarily on broad issues of
phenomenology, the hidden sector fields we will consider are
the dilaton  $S$ in the chiral multiplet formulation
and the three diagonal K\"ahler moduli $\supal{T}$.

In the orbifold compactifications we study here, the
tree-level K\"ahler potential for the moduli and the matter
superfields is known. For the moduli sector we have
\begin{equation}
K(S,\oline{S};\supal{T},\oline{T}^{\alpha})  = k(S+\bar S) -
\sum_{\alpha=1}^3 \ln \(T^\alpha + \oline{T}^\alpha \) ,
\label{Kmod} \end{equation}
where we prefer to leave the form of the dilaton K\"ahler
potential $k$ unspecified at this point. Its precise form depends
on how one stabilizes the dilaton. We will return to this issue
when we discuss the dilaton-dominated scenario of
Section~\ref{sec:dilaton}. As for the observable sector matter
fields $Z^{i}$ with modular weights $n_{i}^{\alpha}$ associated
with each of the three $T^{\alpha}$, we assume a diagonal K\"ahler
metric given by
\begin{equation}
K_{i \bar{j}} = \kappa_i(Z^n) \delta_{ij} + O(|Z^i|^2),
\end{equation}
with
\begin{equation}
\kappa_i(Z^n) = \prod_\alpha (T^\alpha +
\oline{T}^\alpha)^{n_{i}^{\alpha}} .
\end{equation}
In the interests of simplicity, we assume that
the three K\"ahler moduli $T^{\alpha}$ can be treated as
equivalent so that
\begin{equation}
K(S,\oline{S};T,\oline{T}) = k(S+\oline{S})-3\ln(T+\oline{T});
\quad \quad \kappa_{i} =(T+\oline{T})^{n_{i}},
\label{kappa} \end{equation}
where $n_{i} = \sum_{\alpha}n_{i}^{\alpha}$.

The tree-level gauge kinetic functions $f_{a}(Z^n)$, one for each
gauge group ${\cal G}_{a}$, are given in the weak coupling regime
by
\begin{equation}
f_{a}^{0}(Z^n)=S .
\label{ftree} \end{equation}
Their vacuum expectation values give the associated gauge
couplings $\lang \re f_{a} \rang = 1/g_{a}^{2}$. To~(\ref{ftree})
will be added certain string threshold corrections when we exhibit
the one loop soft supersymmetry breaking terms below.

The scalar potential, written in terms of auxiliary fields, is
given by the expression\footnote{We will assume vanishing D-terms
in what follows.}
\begin{equation}
V= K_{I\oline{J}} F^I  \oline{F}^{\bar{J}} - \frac{1}{3} M
\oline{M} ,
\label{pot} \end{equation}
with $K_{I\oline{J}}= \partial^2 K / \partial Z^I \partial
\oline{Z}^{\bar{J}}$ being the K\"ahler metric. The fields $F^I$
in~(\ref{pot}) are the auxiliary fields associated with the chiral
superfields $Z^I$ while the field $M$ is the auxiliary field of
the supergravity multiplet. Solving the equations of motion for
these auxiliary fields yields
\begin{eqnarray}
F^M&=&- e^{K/2} K^{M\oline{N}} \left(\oline{W}_{\oline{N}} +
K_{\oline{N}} \oline{W} \right), \label{Faux} \\
\oline{M}&=& -3e^{K/2}\oline{W}, \label{Maux}
\end{eqnarray}
with $K^{M \oline{N}}$ being the inverse of the K\"ahler metric.
Note that these expressions are given in terms of reduced Planck
mass units where we have set $M_{\PL}/\sqrt{8\pi} = 1$. The
supergravity auxiliary field is related to the gravitino mass by
\begin{equation}
M_{3/2} = -\frac{1}{3}\lang \oline{M} \rang = \lang e^{K/2}
\oline{W} \rang .
\label{gravmass} \end{equation}

We adopt the ansatz of Brignole et al.~\cite{BrIbMu94} in
which one assumes that the communication of supersymmetry breaking
from the hidden sector to the observable sector occurs through the
agency of one of the moduli -- in this case either the dilaton $S$
or the (universal) K\"ahler modulus $T$ -- by the presence of a
non-vanishing vacuum expectation value of their auxiliary fields
$F^{S}$ or $F^{T}$. In principle both types of moduli could
participate in supersymmetry breaking, and so one typically
introduces a ``Goldstino angle'' $\theta$ to parameterize the
degree to which one sector or the other feels the supersymmetry
breaking. While this is a common practice, all known explicit
models of moduli stabilization in the heterotic string context
predict that this ``angle'' will be some integer multiple of
$\pi/2$; that is, only one of the two classes of moduli
participates in supersymmetry breaking.

If these are the {\em only} sectors with non-vanishing auxiliary
fields in the vacuum, then the further requirement that the
overall vacuum energy at the minimum of the potential~(\ref{pot})
be zero allows us to immediately identify (up to phases, which we
set to zero in what follows)
\begin{eqnarray}
F^S &=& -\frac{1}{\sqrt{3}}\oline{M} k_{s\bar{s}}^{-1/2}
\sin\theta = \sqrt{3}M_{3/2}k_{s\bar{s}}^{-1/2} \sin\theta ,
\label{sin} \\
F^T &=& -\frac{1}{\sqrt{3}}\oline{M} K_{t\bar{t}}^{-1/2}
\cos\theta = \sqrt{3}M_{3/2}K_{t\bar{t}}^{-1/2} \cos\theta ,
\label{cos}
\end{eqnarray}
where the last equality holds for the vacuum expectation values,
using~(\ref{gravmass}) above. We should note that the condition of
vanishing vacuum energy is a necessary one to employ the above
parameterization. In this work we will avoid discussing specific
models of dynamical supersymmetry breaking and moduli
stabilization, but any such model must include some mechanism for
engineering a vanishing vacuum energy in order to make contact
with the results presented here.

\subsection{Modular invariance and tree level supersymmetry
breaking terms}

The soft supersymmetry breaking terms in string-derived
supergravity depend on the moduli through the observable sector
superpotential and this is, in turn, determined by insisting on
modular invariance of the low-energy effective Lagrangian. The
diagonal modular transformations
\begin{equation}
T \to \frac{aT - ib}{icT +d}, \quad ad-bc=1, \; \;a,b,c,d \in Z ,
\label{modtrans} \end{equation}
leave the classical effective supergravity theory invariant,
though at the quantum level these transformations are
anomalous~\cite{DiKaLo91,AnNaTa91,CaOv93,DeFeKoZw92}. This anomaly
is cancelled in the effective theory by the presence of a
universal Green-Schwarz counterterm and model-dependent string
threshold corrections~\cite{GaTa92,KaLo95}, which we describe
below.

A matter field $Z^{i}$ of modular weight $n_{i}$ transforms
under~(\ref{modtrans}) as
\begin{equation} Z^i \to
(icT + d)^{n_i} Z^i \label{mattertrans}
\end{equation}
while the K\"ahler potential of~(\ref{kappa}) undergoes a K\"ahler
transformation \mbox{$K \to K + 3(F+\oline{F})$}, with
$F=\ln(icT+d)$, under~(\ref{modtrans}). Therefore the classical
symmetry will be preserved provided the superpotential transforms
as~\cite{BiGiGr01}
\begin{equation}
W \to W \(icT + d\)^{-3} . \label{pottrans}
\end{equation}
To ensure this transformation property the superpotential of
string-derived models has a moduli dependence of the form
\begin{equation} W_{ijk} = w_{ijk} \left[
\eta(T) \right]^{-2(3+n_i+n_j+n_k)}. \label{WT}
\end{equation}
where $W_{ijk} = \partial^3W(Z^N)/\partial Z^i\partial Z^j\partial
Z^k$. The function $\eta(T)$ is the classical Dedekind eta
function
\begin{equation}
\eta(T) = e^{-\pi T /12} \prod_{n=1}^{\infty} (1-e^{-2\pi nT})
\end{equation}
and it has a well-defined transformation under~(\ref{modtrans})
given by
\begin{equation}
\eta(T) \to \(icT + d\)^{1/2} \eta(T) .
\end{equation}
We will also have to introduce the modified Eisenstein
function
\begin{equation}
\Eisen \equiv 2\zeta(t) + \frac{1}{t+\bar{t}} , \quad \mathrm{where} \quad
\zeta(T) = \frac{1}{\eta(T)} \frac{d\eta(T)}{dT}
\label{Eisenstein} \end{equation}
which vanishes at the self-dual points $t=1$ and $t=e^{i\pi/6}$.

We are now in a position to give the tree level soft supersymmetry
breaking terms. The tree level gaugino mass for canonically
normalized gaugino fields is simply
\begin{equation}
M^{0}_a =  \frac{g_a^2}{2} F^n \partial_n f_{a}^{0} .
\label{gaugtree} \end{equation}
We define our trilinear A-terms and scalar masses for canonically
normalized fields by
\begin{equation}
V_A = \frac{1}{6}\sum_{ijk}A_{ijk}e^{K/2}W_{ijk}z^i z^j z^k + \hc
= \frac{1}{6}\sum_{ijk}A_{ijk}e^{K/2}(\kappa_i \kappa_j
\kappa_k)^{-1/2}W_{ijk}\h{z}^i \h{z}^j \h{z}^k + \hc,
\label{Apot} \end{equation}
where $\h{z}^i = \kappa_i^{{1/2}}z^i$ is a normalized scalar
field, and by
\begin{equation}
V_M = \sum_i M^2_i\kappa_i|z^i|^2= \sum_i M^2_i|\h{z}^i|^2 .
\end{equation}
With these conventions our tree level expressions are
\begin{eqnarray}
A^{0}_{ijk}&=&\lang F^n\partial_n\ln(\kappa_i\kappa_j\kappa_k
e^{-K}/W_{ijk})\rang. \label{Atree} \\ (M^{0}_{i})^2&=& \lang
\frac{M\oline{M}}{9} - F^n
\oline{F}^{\bar{m}}\partial_n\partial_{\bar{m}}\ln\kappa_i \rang .
\label{scalartree}
\end{eqnarray}
If we specialize now to the case of~(\ref{sin}) and~(\ref{cos})
with moduli dependence given by~(\ref{kappa}),~(\ref{ftree})
and~(\ref{WT}), then the tree level gaugino
masses~(\ref{gaugtree}), A-terms~(\ref{Atree}) and scalar
masses~(\ref{scalartree}) become
\begin{eqnarray}
M_{a}^{0}&=&\frac{g_{a}^{2}}{2}F^{S} \nonumber \\
A_{ijk}^{0} &=& (3+n_i + n_j + n_k)\Eisen F^{T} -k_{S}F^{S}
\nonumber \\
\(M_{i}^{0}\)^{2}&=&\frac{M\oline{M}}{9} + n_{i} \frac{|F^{T}
|^{2}}{(t + \bar{t})^{2}} .
\label{treeBIM} \end{eqnarray}
The expressions in~(\ref{treeBIM}) are to be understood as vacuum
values, though we will drop the cumbersome brackets
$\lang\dots\rang$ from here onwards.

These tree level soft terms have been studied extensively in the
past and we wish to understand what impact the less-studied loop
corrections can have on the phenomenology of orbifold models. We
can now appreciate the importance of the two cases we intend to
study in Sections~\ref{sec:moduli} and~\ref{sec:dilaton}. In the
case where the modular weights of the observable sector are
universally $n_{i} = -1$, as would be the case for the untwisted
sector of orbifold compactifications, then from~(\ref{WT}) it is
clear that the moduli do not couple to the observable sector
through the superpotential and we have a true no-scale model.
Indeed, substituting~(\ref{cos}) into~(\ref{treeBIM}) we find that
if the dilaton does not participate in supersymmetry breaking all ($\theta=0$)
tree level soft terms are precisely zero in such a case. This is
the situation we will investigate in Section~\ref{sec:moduli}.

Alternatively, if the dilaton {\em is} the primary source of
supersymmetry breaking in the observable sector, loop-level
corrections to at least the trilinear A-terms and gaugino masses
will continue to be important provided the magnitude of $F^S$ is
suppressed relative to that of $M$, as it is in models where
nonperturbative corrections to the K\"ahler potential are used to
stabilize the dilaton~\cite{BiGaWu97b}. In such a scenario the
modular weights of the matter fields are irrelevant at the tree
level, so we will continue to assume $n_{i} = -1$ for the sake of
convenience when we investigate these models in
Section~\ref{sec:dilaton}. We thus need to turn our attention to
the study of soft supersymmetry breaking at one loop.

\subsection{General one loop corrections to soft supersymmetry
breaking}

In this section we aim to provide sufficient background to justify
the form of the one loop soft term expressions which are our goal,
as well as explain some notation we will need for our
phenomenological analysis. For more complete explanations of what
is contained in this section, including expressions with arbitrary
modular weights and three independent K\"ahler moduli, the reader
should consult the precursor to this work~\cite{BiGaNe01}.

We begin with gaugino masses which can be understood as a sum of
loop-induced contributions from the field theory point of view, and terms
that can be thought of as one loop stringy corrections. The field
theory loop contribution can be derived completely from the
superconformal anomaly and is given by~\cite{GaNeWu99,BaMoPo00}
\begin{equation}
M^{1}_a|_{\rm an} = \frac{g_{a}^{2}(\mu)}{2} \left[ \frac{2
b_a}{3} \oline{M} - \frac{1}{8\pi^2} \left( C_a - \sum_i C^i_a
\right) F^n K_{n} - \frac{1}{4\pi^2} \sum_i C^i_a F^n \partial_n
\ln \kappa_i \right],
\label{Man} \end{equation}
where $C_a$, $C_a^i$ are the quadratic Casimir operators for the
gauge group ${\cal G}_a$ in the adjoint representation and in the
representation of $Z^i$, respectively, and $b_{a}$ is the
beta-function coefficient for the group ${\cal G}_{a}$:
\begin{equation}
b_a = \frac{1}{16\pi^2} \( 3 C_a - \sum_i C_a^i \) .
\label{ba} \end{equation}

As mentioned in the previous section one expects modular anomaly
cancellation to occur through a universal Green-Schwarz
counterterm with group-independent coefficient $\delta_{\GS}$.
Such a term can be thought of as a loop correction that
contributes to gaugino masses in the form
\begin{equation}
M^{1}_a|_{\GS} = \frac{g^2_a(\mu)}{2}  \frac{2F^T}{(t + \bar{t})}
\frac{\delta_{\GS}}{16\pi^2} .
\label{MGS} \end{equation}
In addition string threshold corrections generally appear in the
effective theory, which may be interpreted as one loop corrections
to the gauge kinetic functions of the form
\begin{equation}
f^{1}_{a}(Z^n)=\ln \eta^2(T) \[\frac{\delta_{\GS}}{16\pi^2}+b_a\],
\label{floop} \end{equation}
which generate loop contributions to gaugino masses given by
\begin{equation}
M^{1}_a|_{\rm th} = \frac{g^{2}_{a}(\mu)}{2}
\[\frac{\delta_{\GS}}{16\pi^2}+ b_a\]4 \zeta(t)F^{T} .
\label{Mth}\end{equation}
Putting together the tree level gaugino mass with the loop
contributions~(\ref{Man}),~(\ref{MGS}) and~(\ref{Mth}) gives
\begin{equation}
M_{a} = \frac{g_{a}^{2}\(\mu\)}{2} \lbr 2
 \[ \frac{\delta_{\GS}}{16\pi^{2}} + b_{a}
\]\Eisen F^{T} + \frac{2}{3}b_{a}\oline{M} +\[ 1
- 2 b_{a}' k_s \] F^{S} \rbr \label{Maloop}
\end{equation}
where we have defined the quantity
\begin{equation}
b_{a}' = \frac{1}{16\pi^2} \(C_a - \sum_i C_a^i \) .
\label{baprime}
\end{equation}

To understand the form of the one-loop A-terms and scalar masses
it is necessary to describe how field theory loops
are regulated in supergravity, seen as an effective theory of strings.
The regulation of matter and Yang-Mills loop
contributions to the matter wave function renormalization requires
the introduction of Pauli-Villars chiral superfields $\Phi^A =
\Phi^{i}$, $\wh{\Phi}^{i}$ and $\Phi^a$ which transform according
to the chiral matter, anti-chiral matter and adjoint
representations of the gauge group and have signatures $\eta_A =
-1,+1,+1,$ respectively.  These fields are coupled to the light
fields $Z^{i}$ through the superpotential
\begin{equation}
W(\Phi^A,Z^i) = {1\over2}W_{ij}(Z^k)\Phi^i \Phi^j + \sqrt{2}\Phi^a
\wh{\Phi}_{i} (T_a Z)^i + \cdots, \label{PVcoup}
\end{equation}
where $T_a$ is a generator of the gauge group, and their K\"ahler
potential takes the schematic form
\begin{equation}
K_{\PV} = \sum_A \kappa_A^{\Phi}(Z^{N})|\Phi^A|^2, \label{PVKahler}
\end{equation}
where the functions $\kappa_{A}$ are {\em a priori} functions of
the hidden sector (moduli) fields. These regulator fields must be
introduced in such a way as to cancel the quadratic divergences of
the light field loops -- and thus their K\"ahler potential is
determined relative to that of the fields which they regulate.

The PV mass for each superfield $\Phi^A$ is generated by coupling
it to another field $\Pi^{A}=(\Pi^i, \widehat{\Pi}^i, \Pi^a)$ in
the representation of the gauge group conjugate to that of
$\Phi^A$ through a superpotential term
\begin{equation}
W_m = \sum_{A}\mu_{A}(Z^N)\Phi^A \Pi^A, \label{PVbilinear}
\end{equation}
where $\mu_{A}(Z^N)$ can in general be a holomorphic function of
the light superfields. There is no constraint on the K\"ahler
potential for the fields $\Pi^A$ as there was for those of the
$\Phi^A$. However, if we demand that our regularization preserve
modular invariance then we can determine the moduli dependence of
the regulator fields $\Phi^A$
\begin{equation}
\lbr \begin{array}{l} \Phi^i:\;\;\kappa^\Phi_i = \kappa_i = (T +
\oline{T})^{n_i}, \\ \wh{\Phi}^i:\;\; \h{\kappa}^\Phi_i =
\kappa^{-1}_i, \\ \Phi^a: \;\;\kappa^\Phi_a = g_a^{-2}e^K =
g_a^{-2}e^k(T + \oline{T})^{-3},
\end{array} \right.
\label{kapPhi}
\end{equation}
and we can determine the moduli dependence of the supersymmetric
mass $\mu_{A}(Z^N)$ in~(\ref{PVbilinear})
\begin{equation}
\mu_{A}(Z^{N}) = \mu_{A}(S)
\[\eta(T)\]^{-2(3+n_A+ q_A)},
\end{equation}
with $n_A$ and $q_A$ being the modular weights of the fields
$\Phi^A$ and $\Pi^A$, respectively. Furthermore, we can postulate
the form of the moduli dependence of $\kappa_A$ for the
mass-generating fields
\begin{equation}
\Pi^A:\;\;\kappa^\Pi_A = h_A(S+\oline{S})(T + \oline{T})^{q_A}.
\label{kapPi}
\end{equation}
So at this point the dilaton dependence in the superpotential
term~(\ref{PVbilinear}) and the functions $h_A$, as well as the
modular weights $q_A$ of the fields $\Pi^A$, are new free
parameters of the theory introduced at one loop as a consequence
of how the theory is regulated. Given~(\ref{PVbilinear}) we can
extract the Pauli-Villars masses that appear as regulator masses
in the logarithms at one loop
\begin{equation}
m_A^2 = e^K (\kappa^\Phi_A)^{-1/2} (\kappa^{\Pi}_A)^{-1/2} |\mu_A|^2 ,
\label{PVmass}
\end{equation}
with $m_A = (m_i, \h{m}_i, m_a)$ being the masses of the regulator
fields $\Phi^i, \wh{\Phi}^{i}, \Phi^a$, respectively.

In terms of these regulator masses, the complete one loop
correction to the trilinear A-terms and scalar masses in a general
supergravity theory was given in~\cite{GaNe00b}. In this survey we
will simplify the expressions and reduce the parameter space by
making some reasonable assumptions. Let us first assume that the
functions $\mu_{A}(Z^N)$ that appear in~(\ref{PVbilinear})
and~(\ref{PVmass}) are proportional to one overall Pauli-Villars
scale $\mu_{\PV}$ so that $\h{\mu}_{i} = \mu_{a} = \mu_{i} \equiv
\mu_{\PV}$. This scale is presumed to represent some fundamental
scale in the underlying string theory. Let us further assume that there
is no dilaton dependence of the PV masses so that
$h_A(S+\oline{S})$ is trivial and $\mu_{\PV}$ is constant. With
these simplifications the complete trilinear A-term at one loop is
given by
\begin{eqnarray}
A_{ijk} &=& \frac{1}{3} A^{0}_{ijk} - \frac{1}{3}
\gamma_{i}\oline{M} - \Eisen F^T \(\sum_a\gamma_i^a p_{ia} +
\sum_{lm}\gamma_i^{lm}p_{lm}\) \nonumber \\
 & & -\ln\[(t + \bar{t})|\eta(t)|^4\]\(2
\sum_a\gamma_i^a p_{ia}M^{0}_a +
\sum_{lm}\gamma_i^{lm}p_{lm}A^{0}_{ilm}\) \nonumber \\ & & +
2\sum_a\gamma_i^a M^{0}_a \ln(\mu_{\PV}^{2}/\mu_R^2) +
\sum_{lm}\gamma_i^{lm}A^{0}_{ilm}\ln(\mu_{\PV}^{2}/\mu_R^2)
 + {\rm cyclic}(ijk) ,
\label{Atermraw}
\end{eqnarray}
where we have defined the following combinations of modular
weights from the Pauli-Villars sector
\begin{equation}
p_{ij} = 3 + \frac{1}{2} \(n_{i} + n_j + q_i + q_j\), \quad p_{ia}
= \frac{1}{2} \(3 + q_{a} +\h{q}_{i} - n_{i} \)
\label{PVmodsraw} \end{equation}
which we will refer to as ``regularization weights'' in reference
to their origin from the PV sector of the theory.\footnote{Note
that as we are considering the case with universal modular weights
$n_i = -1$, these expressions can be reduced to $p_{ij} =
2+\frac{1}{2}(q_i + q_j)$ and $p_{ia} = 2 +\frac{1}{2}(q_a +
\h{q}_i)$.}

In~(\ref{Atermraw}) $M^{0}_a$ and $A^{0}_{ilm}$ are the tree level
gaugino masses and A-terms given in~(\ref{treeBIM}) and the
parameters $\gamma$ determine the chiral multiplet wave function
renormalization
\begin{eqnarray}
\gamma^j_i = \frac{1}{32\pi^2}\[4\delta^j_i\sum_a g^2_a(T^2_a)^i_i
- e^K\sum_{kl}W_{ikl}\oline{W}^{jkl}\] . \label{gam}
\end{eqnarray}
We have implicitly made the approximation that generational mixing
is unimportant and can be neglected in~(\ref{Atermraw}), and that
motivates the definitions
\begin{eqnarray}
\gamma_i^j &\approx&\gamma_i\delta^j_i, \quad \gamma_i =
\sum_{jk}\gamma_i^{jk} + \sum_a\gamma_i^a, \nonumber \\ \gamma_i^a
&=& \frac{g^2_a}{8\pi^2}(T^2_a)^i_i, \quad \gamma_i^{jk} =
-\frac{e^K}{32\pi^2}
(\kappa_i\kappa_j\kappa_k)^{-1}\left|W_{ijk}\right|^2.\label{diag}
\end{eqnarray}
The scalar masses are obtained similarly and take the form
\begin{eqnarray}
\(M_{i}\)^{2} &=& (M_{i}^{0})^{2}+ \gamma_{i}\frac{M\oline{M}}{9}
- \frac{|F^{T}|^{2}}{(t + \bar{t})^{2}} \(\sum_a\gamma^a_i p_{ai}
+ \sum_{jk}\gamma^{jk}_i p_{jk}\) \nonumber \\ & & +\lbr
\frac{M}{3} \[\sum_a\gamma^a_i M^{0}_{a} +
{1\over2}\sum_{jk}\gamma^{jk}_iA^{0}_{ijk}\] + \hc \rbr \nonumber
\\ & & + \lbr  F^T \Eisen
\(\sum_a\gamma_i^a p_{ia}M^{0}_a +
\frac{1}{2}\sum_{jk}\gamma_i^{jk}p_{jk} A^{0}_{jk}\) + \hc
\rbr\nonumber \\ & & - \ln\[(t +
\bar{t})|\eta(t)|^4\]\lbr\sum_a\gamma_i^a p_{ia}
\[3(M^{0}_a)^2 - (M^{0}_i)^2\]\right. \nonumber \\
& &  \qquad\qquad +
\left.\sum_{jk}\gamma_i^{jk}p_{jk}\[(M^{0}_j)^2 + (M^{0}_k)^2 +
(A^{0}_{ijk})^2\]\rbr \nonumber \\ & & +
\sum_a\gamma_i^a\[3(M^{0}_a)^2 - (M^{0}_i)^2\]
\ln(\mu_{\PV}^{2}/\mu_R^2) \nonumber \\ & & +
\sum_{jk}\gamma_i^{jk}\[(M^{0}_j)^2 + (M^{0}_k)^2 +
(A^{0}_{ijk})^2\]\ln(\mu_{\PV}^{2}/\mu_R^2), \label{massraw}
\end{eqnarray}
with $M_i^{0}$ being the tree level scalar masses
of~(\ref{treeBIM}).

To put these expressions into a less cumbersome form we will
consider the case where the various regularization weights
$p_{ia}$ and $p_{jk}$ can be treated as one overall parameter $p$.
Then inserting the tree level soft terms~(\ref{treeBIM})
into~(\ref{Atermraw}) and~(\ref{massraw}) yields these soft terms
in the final form we will use in the following analysis:
\begin{eqnarray}
A_{ijk} &=& -\frac{k_s}{3}F^S - \frac{1}{3} \gamma_{i}\oline{M} -
p \gamma_{i} \Eisen F^{T} + \tilde{\gamma}_{i} F^{S} \lbr
\ln(\mu_{\PV}^{2}/\mu_R^2) -p\ln\[(t+\bar{t}) |\eta(t)|^4\] \rbr
\nonumber \\ & & + {\rm cyclic}(ijk) \label{finalA} \\ M_{i}^{2}
&=& \lbr \frac{|M|^2}{9} -\frac{|F^T|^{2}}{(t+\bar{t})^{2}}\rbr \[
1 + p\gamma_i -\(\sum_{a}\gamma_{i}^{a}
-2\sum_{jk}\gamma_{i}^{jk}\) \( \ln(\mu_{\PV}^{2}/\mu_R^2)
-p\ln\[(t+\bar{t}) |\eta(t)|^4\] \)
\] \nonumber \\
 & &+ (1-p)\gamma_i \frac{|M|^2}{9} + \lbr
 \wtd{\gamma}_{i}\frac{MF^S}{6}+\hc \rbr +\lbr p \wtd{\gamma}_{i}\Eisen
 \frac{\oline{F}^{T}F^S}{2} + \hc \rbr \nonumber \\
 & & +|F^{S}|^2 \[ \(\frac{3}{4}\sum_{a} \gamma_{i}^{a} g_{a}^{4}
 + k_s k_{\bar{s}} \sum_{jk} \gamma_{i}^{jk}\) \(
\ln(\mu_{\PV}^{2}/\mu_R^2) -p\ln\[(t+\bar{t}) |\eta(t)|^4\] \) \],
\label{finalscalar}
\end{eqnarray}
where $\wtd{\gamma}_{i}$ is a shorthand notation for
\begin{equation}
\wtd{\gamma}_{i} = \sum_{a} \gamma_{i}^{a} g_{a}^{2} - k_{s}
\sum_{jk} \gamma_{i}^{jk} . \label{tildegamma}
\end{equation}
The adoption of one overall regularization weight $p$ makes it
possible to identify the quantity $\ln(\mu_{\PV}^{2}/\mu_R^2)
-p\ln\[(t+\bar{t}) |\eta(t)|^4\]$ as a stringy threshold
correction to the overall PV mass scale, or effective cut-off,
$\mu_{\PV}$. The one loop expressions
of~(\ref{Maloop}),~(\ref{finalA}) and~(\ref{finalscalar}) will be
our starting point in the following two sections where we
investigate their phenomenological significance in scenarios where
tree level soft supersymmetry breaking is suppressed or vanishing.

\section{Classification of heterotic orbifold models}
\label{sec:space}

As a result of the assumptions made in Section~\ref{sec:general}
our general parameter space is defined by seven quantities, if we
assume a tree-level K\"ahler potential for the dilaton of the form
$k(S,\oline{S})=-\ln(S+\oline{S})$ with $\lang s+\bar{s} \rang
=2/g_{\STR}^{2}\simeq 4$. Three of these parameters represent
scales: the gravitino mass $M_{3/2}$, the boundary condition scale
$\mu_{\UV}$ at which the soft supersymmetry breaking terms appear
and the Pauli-Villars cutoff scale $\mu_{\PV}$. The other four are
parameters describing the moduli sector: the vacuum expectation
value of the real part of the (uniform) K\"ahler modulus $\lang
t+\bar{t}\rang$, the modular weights of the PV regulators
parameterized by $p$, the value of the Green-Schwarz coefficient
$\delta_{\GS}$ and the Goldstino angle $\theta$ which determines
the degree to which the dilaton and moduli F-terms participate in
the transmission of supersymmetry breaking.

\subsection{Moduli dominated scenarios}
\label{sec:moduli}

This section assumes that i) all fields have modular weight
$n_i=-1$, which gives a no--scale structure, and ii) we have
moduli-dominated supersymmetry breaking ($\theta$ vanishes,
$\cos\theta=1$) so that the tree level soft terms are precisely
zero. In addition, those loop-induced soft terms
in~(\ref{finalscalar}) proportional to tree level quantities are
also zero, leaving only those arising from the superconformal
anomaly and those related to the Pauli-Villars sector:
\begin{eqnarray}
M_a &=& \frac{g_{a}^{2}\(\mu\)}{2} \lbr 2
 \[ \frac{\delta_{\GS}}{16\pi^{2}} + b_{a}
\]\Eisen F^{T} + \frac{2}{3}b_{a}\oline{M} \rbr, \nonumber \\
A_{ijk}&=& - \frac{1}{3} \gamma_{i}\oline{M} - p \gamma_{i} \Eisen
F^{T} + {\rm cyclic}(ijk), \nonumber \\ M_{i}^{2} &=& (1-p)\gamma_i
\frac{|M|^2}{9}. \label{modsoft}
\end{eqnarray}

Note that in the special case where the moduli are stabilized at
one of their two self-dual points $t=1$ and $t=e^{i\pi/6}$ we
recover the universal ``anomaly-mediated'' results
\begin{eqnarray}
M_{a}&=&g_{a}^{2}\(\mu\) b_{a}\frac{\oline{M}}{3} \nonumber \\
A_{ijk}&=&-
\(\gamma_{i}+\gamma_{j}+\gamma_{k}\)\frac{\oline{M}}{3} \nonumber
\\ M_{i}^{2}&=&(1-p)\gamma_{i}\frac{|M|^{2}}{9} . \label{amsb}
\end{eqnarray}
The scenario that has come to be referred to as the
``Anomaly-Mediated Supersymmetry Breaking'' (AMSB)
scenario~\cite{GiLuMuRa98,RaSu99,PoRa99}, in which scalar masses
first appear only at two loops, is obtained in the limit as the
phenomenological parameter $p$ approaches unity. This would be the
case, for example, if the mass-generating Pauli-Villars fields
$\Pi^A$ had the same K\"ahler metric as the regulator fields
$\Phi^A$ and thus $q_i = -1$, $\h{q}_{i} =1$ and $q_a = -3$. In
any event, it is clear that the parameter $p$ defines a {\em
family} of ``anomaly mediated'' models for the case of moduli
stabilized at self-dual points. Importantly, for $p<1$, as would
be typical in orbifold models, the anomaly-mediated masses for the
matter scalars are positive with only the Higgs fields having
negative squared masses at the scale $\mu_{\UV}$.

The contrast between these results and those of the standard AMSB
case was discussed in~\cite{GaNe00b,BiGaNe01}, but it is important
to point out that the vanishing of the scalar masses at one loop
in the $p\to 1$ limit clearly demands a two-loop analysis of these
soft terms. A treatment of supergravity radiative corrections,
regularized by the Pauli-Villars mechanism, at two loops would be
a massive undertaking. Some subset of these corrections would
undoubtedly be the soft terms found in~\cite{RaSu99,PoRa99} using
a spurion technique. As is now well known, these two-loop terms
arising from the superconformal anomaly would imply negative
squared masses for the scalar leptons. To remedy this situation,
the AMSB model, as it has been institutionalized in studies such
as the Snowmass Points and Slopes~\cite{snowmass}, adopts an
otherwise {\em ad hoc} scalar mass contribution that is universal
and sufficiently large. Not surprisingly, these models are often
characterized as having light sleptons as a key feature of the
phenomenology.

While several extensions of the original (or ``minimal'') AMSB
model now exist which address the slepton mass problem in a
variety of ways,\footnote{For some early solutions,
see~\cite{AMSB}.} the generalized anomaly mediated model we
present here suffers from no such problem when $p \neq 1$. While
the phenomenology associated with the gaugino mass spectrum is
familiar from past studies of the AMSB paradigm, the scalar mass
sector looks quite different. We will begin our analysis in
Section~\ref{sec:AMSB} with this generalized anomaly mediated
model (which we will refer to as the ``PV--AMSB model'') before
enlarging the parameter space to include all moduli-dominated
models in Sections~\ref{sec:modplots} and~\ref{sec:gsplots}.

\subsection{Dilaton dominated scenarios with K\"ahler suppression}
\label{sec:dilaton}

We next turn to the opposite extreme, where
the dilaton is the primary source of supersymmetry breaking in the
observable sector ($\sin\theta =1$). In such a scenario we would
ordinarily expect the loop corrections such as those
in~(\ref{Man}),~(\ref{Atermraw}) and~(\ref{massraw}) to be small
perturbations on the dominant (and universal) tree level
contributions and thus safely neglected. The phenomenology of the
dilaton-domination scenario at tree level has been extensively
studied in the literature~\cite{dildom}.

The bulk of these studies, however, have considered only the case
of the standard K\"ahler potential
$k(S,\oline{S})=-\ln(S+\oline{S})$ derived from the tree level
string theory. It is well known that when this form is used,
stabilizing the dilaton at acceptable weak-coupling values is
extremely difficult. In the case where a dilaton potential is
imagined to arise through field theory nonperturbative effects
(such as gaugino condensation) it has been shown that one
condensate alone can never succeed in stabilizing the dilaton in
such a way as to reproduce the weak coupling observed in
nature~\cite{BaDi94,Ca96}. Furthermore, while multiple condensates
can be used to provide the requisite
stabilization~\cite{CaLaMuRo90,deCaMu93}, these models tend to
predict a moduli-dominated scenario such as those in
Section~\ref{sec:moduli}.

In contrast, if one postulates a nonperturbative correction of
stringy origin to the dilaton K\"ahler potential, as was first
motivated by Shenker~\cite{Sh90}, then one condensate can indeed
stabilize the dilaton at weak coupling while simultaneously
ensuring vanishing vacuum energy at the minimum of the potential.
The power of this approach was first demonstrated in an explicit
model of modular invariant gaugino condensation
in~\cite{BiGaWu96,BiGaWu97a} and confirmed in~\cite{BadeCo98},
providing a concrete realization of the so-called ``generalized
dilaton domination scenario''~\cite{Ca96}.

The key feature of such models is the deviation of the dilaton
K\"ahler metric from its tree level value. If we imagine the
superpotential for the dilaton having the form $W(S) \propto
e^{-3S/2b_{+}}$, with $b_{+}$ being the largest beta-function
coefficient~(\ref{ba}) among the condensing gauge groups
of the hidden sector, then it is clear
from~(\ref{Faux}) and~(\ref{Maux}) that requiring the
potential~(\ref{pot}) to vanish implies
\begin{equation}
(k^{s\bar{s}})\left|k_s - \frac{3}{2b_{+}} \right|^{2} =3 \; \;
\to (k^{s\bar{s}})^{-1/2} = \sqrt{3}
\frac{\frac{2}{3}b_{+}}{1-\frac{2}{3}b_{+}k_{s}} . \label{Ktrue}
\end{equation}
The condition in~(\ref{Ktrue}) is independent of the means by
which the dilaton is stabilized and is a result merely of
requiring a vanishing vacuum energy in the dilaton-dominated
limit. The explicit model of~\cite{BiGaWu97a}, however, was able
to achieve precisely this relation with $\lang k_s \rang =
-g_{\STR}^{2}/2$, using a correction to the dilaton action that
involved tuning $\order(1)$ numbers only.\footnote{In fact, the
model considered in this reference used the linear multiplet
formalism for the dilaton in which such nonperturbative
corrections are more easily incorporated into the low energy
effective supergravity theory. For a description on how to
correctly translate from one formulation to the other, see
Appendix A of~\cite{BiGaNe01}.}

We can parameterize the departure that~(\ref{Ktrue}) represents
from the tree level K\"ahler metric $\lang (k_{s\bar{s}}^{\rm
tree})^{1/2} \rang = \lang 1/(s+\bar{s}) \rang = g_{\STR}^{2}/2
\simeq 1/4$ by introducing the phenomenological parameter
\begin{equation}
a_{\rm np} \equiv \(\frac{k_{s\bar{s}}^{\rm
tree}}{k_{s\bar{s}}^{\rm true}}\)^{1/2} \label{acond}
\end{equation}
so that the auxiliary field of the dilaton chiral supermultiplet
can be expressed as
\begin{equation}
F^{S} = \sqrt{3} M_{3/2} (k_{s\bar{s}})^{-1/2} =
\sqrt{3}M_{3/2}a_{\rm np}(k_{s\bar{s}}^{\rm tree})^{-1/2} .
\label{FS}
\end{equation}
The importance of~(\ref{Ktrue}) for our purposes is to recognize
that the factor of $b_{+}$, containing as it does a loop factor,
will suppress the magnitude of the auxiliary field $F^S$ relative
to that of the supergravity auxiliary field $M$ through the
relation~(\ref{FS}). Therefore tree level soft terms for the
gaugino masses and trilinear A-terms will be of comparable
magnitude to the loop-induced soft terms arising from the
superconformal anomaly (though scalar masses will be only
negligibly altered from their tree level values). We are thus led
to consider the phenomenology of models given by the following
pattern of soft supersymmetry breaking terms:
\begin{eqnarray}
M_{a}&=&\frac{g_{a}^{2}\(\mu\)}{2} \lbr  \frac{2}{3}b_{a}\oline{M}
+\[ 1 - 2 b_{a}' k_s \] F^{S} \rbr \nonumber \\ A_{ijk} &=&
-\frac{k_s}{3}F^S - \frac{1}{3} \gamma_{i}\oline{M} +
\tilde{\gamma}_{i} F^{S} \lbr \ln(\mu_{\PV}^{2}/\mu_R^2)
-p\ln\[(t+\bar{t}) |\eta(t)|^4\] \rbr + {\rm cyclic}(ijk)
\nonumber \\ M_{i}^{2} &=& \frac{|M|^2}{9}
 \[ 1 + \gamma_i
-\(\sum_{a}\gamma_{i}^{a} -2\sum_{jk}\gamma_{i}^{jk}\) \(
\ln(\mu_{\PV}^{2}/\mu_R^2) -p\ln\[(t+\bar{t}) |\eta(t)|^4\] \) \]
\nonumber \\
 & &+ \lbr
 \wtd{\gamma}_{i}\frac{MF^S}{6}+\hc \rbr , \label{BGWsoft}
\end{eqnarray}
where we have dropped terms of $\order\(1/(16\pi^2)^3\)$ in the
scalar masses. These models will be studied in
Section~\ref{sec:dilplots}.

\section{Experimental and cosmological constraints}
\label{sec:analysis}

In order to translate the spectra defined by the soft
supersymmetry breaking terms in Sections~\ref{sec:moduli}
and~\ref{sec:dilaton}, a low energy soft Lagrangian is obtained by
solving the renormalization group equations (RGEs) from some
initial boundary condition scale to the electroweak scale. Having
thus extracted the soft supersymmetry breaking parameters at the low-energy
scale, the physical mass spectra for the superpartner and Higgs
sector can be obtained and various direct collider constraints and
indirect non-collider constraints can be applied to examine the
region of phenomenological viability of these models. In this
section we describe the tools we use and the
various observational constraints we impose.

\subsection{Electroweak symmetry breaking and physical masses}
\label{subsect:elect}

We have used the Fortran code {\tt SuSpect}~\cite{Suspect} to solve the  RGEs for the
soft supersymmetry breaking parameters
between the initial high energy scale
$\mu_{\UV}$ and the scale given by the Z-boson mass. While the
initial scale $\mu_{\UV}$ should itself be treated as a
model-dependent parameter, we have chosen
$\mu_{\UV} = \mu_{\GUT}$ throughout. We use $\tan\beta$ and the
sign of the supersymmetric $\mu$ parameter in the superpotential
as free parameters, defined at the low-energy (electroweak) scale.

The magnitude of the $\mu$ parameter is determined by imposing
electroweak symmetry breaking (EWSB) at the scale defined by the
geometric mean of the two stop masses $(m_{\tilde t_1} m_{\tilde
t_2})^{1/2}$, which minimizes the one-loop scalar effective
potential~\cite{GaRiZw90,deCa93}. The one-loop corrected $\mu$
term is obtained from the condition
\begin{equation}
{\bar{\mu}}^{2}=\frac{\(m_{H_d}^{2}+\delta m_{H_d}^{2}\) -
  \(m_{H_u}^{2}+\delta m_{H_u}^{2}\) \tan{\beta}}{\tan^{2}{\beta}-1}
-\frac{1}{2} M_{Z}^{2} ,
\label{radmuterm} \end{equation}
where $\delta m_{H_u}^2$ and $\delta m_{H_d}^2$ represent the one
loop tadpole corrections to the running Higgs masses $m_{H_u}^2$
and $m_{H_d}^2$~\cite{ArNa92,BaBeOh94,PiBaMaZh97}. {\tt SuSpect}
includes the corrections from all the third generation fermion and
sfermion loops as well as loops from gauge bosons, Higgs bosons,
charginos and neutralinos. As the superpartner spectrum depends on
the value of the $\mu$ parameter computed in~(\ref{radmuterm}),
the proper inclusion of superpartner thresholds in the RG
evolution requires that the procedure be iterated until a coherent
and stable value for $\mu$ is obtained. Usually, {\tt SuSpect}
requires only two or three iterations to have a relative precision
of the order of $10^{-4}$. We have used three.
 The soft supersymmetry breaking parameters at the weak
scale are then passed to the C code {\tt
micrOMEGAs}~\cite{Micromegas} to perform the calculation of
physical masses for the superpartners and various
indirect constraints, to be described in the following section.

The specific algorithm we employ can be described as follows. The
one-loop order SUSY breaking parameters of the heterotic orbifold
models obtained in the previous section are entered into {\tt
SuSpect} at the high energy scale. Other parameters, including
Standard Model fermion masses and gauge weak couplings as well as
certain EWSB parameters such as $\tan\beta$ are input at the low
energy scale. {\tt SuSpect} runs the renormalization group
evolution of the parameters back and forth between the low energy
scales such as $M_Z$ and the electroweak symmetry breaking scale,
and the high--energy scale such as the GUT scale. This is the case
for the soft SUSY--breaking terms (scalar and gaugino masses,
bilinear and trilinear couplings and tan $\beta$) and $\mu$. This
procedure has to be iterated in order to include SUSY threshold
effects or radiative corrections due to Higgs and SUSY particles.
In the first step, these thresholds are only guessed since the
spectrum has not been calculated yet, and the radiative
corrections are not implemented. The thresholds are properly taken
into account in subsequent iterations.  At the electroweak scale,
the consistency in the calculation of the $\mu$ term is checked
using the expression in~(\ref{radmuterm}). We then call {\tt
micrOMEGAs}, which calculates the SUSY spectrum at one loop order
using {\tt FeynHiggs} for the Higgs sector. {\tt micrOMEGAs} also
calculates the $b \rightarrow s \gamma$ branching ratio as well as
the relic density of the lightest neutralino and the anomalous
magnetic moment of the muon. We have checked the consistency
between the two spectra generated by {\tt SuSpect} and by {\tt
micrOMEGAs}: the maximum deviation was of the order of 4\% and was
confined to specific regimes such as high $\tan\beta$.

The first condition that we require on a given set of soft
supersymmetry breaking masses is that an appropriate vacuum state
appears -- in particular that EWSB occurs. Clearly it is possible
for the value of ${\overline{\mu}}^2$ in~(\ref{radmuterm}) to become negative
for a particular choice of model parameters. In such a situation
EWSB does not occur. The precise domain of parameter space
where this occurs is sensitive to the values of the Yukawa
couplings -- and hence to the choice of quark pole masses assumed.
This makes the SUSY corrections to the bottom quark mass
quite important. We have performed the analysis using the
heavy fermion masses in {\tt SuSpect} of
\begin{equation}
M_t=175.0\GeV,\qquad M_b=4.62\GeV, \qquad M_\tau=1.778\GeV  ,
\label{SUSPECTferm}
\end{equation}
and gauge couplings at the Z-mass of
\begin{equation}
\alpha_{\EM}^{\oline{MS}}(M_Z)=1/127.938,\qquad
\alpha_s^{\oline{MS}}(M_Z)=0.118,\qquad\bar{s}^2_W=0.23117  .
\label{SUSPECTalpha}
\end{equation}
We also reject all points in the parameter space which give a
tachyonic Higgs boson mass (in particular, where $m_A^2 < 0$) or
negative sfermion squared masses.

We next require that the lightest supersymmetric particle (LSP) be
neutral, and thus also reject all model parameters where the
lightest neutralino $\wtd{\chi}^0_1$ is not the LSP. In most of
the rejected cases it is the lightest stau $\wtd{\tau}_1$ or the
gluino $\tilde g$ which becomes lighter than the $\wtd{\chi}^0_1$.
The remaining parameter space is further reduced by limits on
superpartner and Higgs masses from various collider experiments.

Concerning the mass bounds we use, we take the
most recent results combined by  
the LEP Working Group~ \cite{LHWG01,LHWG02}. For
the light CP-even neutral Higgs mass ($m_h$), we
 assume that a 95\% confidence level (CL) lower limit
on $m_h$ is set at 111.5 GeV. The search for an invisibly decaying
Higgs boson in $hZ$ production has allowed a 95\% CL lower limit
on $m_h$ to be set at 114.4 GeV, assuming a production cross
section equal to that in the Standard Model and a 100\% branching
fraction to invisible decays~\cite{Higgslimit}. We believe the
value of 113.5 GeV  serves as a good reference point, and this is
the value we use in our analysis. Concerning the chargino limit,
we take 103.5 GeV, bearing in mind that in some degenerate cases
and for light sleptons the limit can go down to 88
GeV~\cite{charginolimit}. For the squark sector the limit of 97
GeV~\cite{Stoplimit} is implemented. For all mass bounds we should
keep in mind that experimental limits are always given in the
context of a particular SUSY model which is not generally a string
motivated one. The bounds we use could possibly be weakened in
some cases.

\subsection{Indirect constraints on supersymmetric spectra}

Various non-collider observations can be used to further reduce
the allowed parameter space of these loop-dominated orbifold
models. We will focus our attention on the three such processes
that yield the most stringent constraints: the density of relic
neutralino LSPs, the branching ratio for decays involving the
process $b \to s \gamma$ and the measurement of the anomalous
magnetic moment of the muon.

\subsubsection{The relic neutralino density}

The existence of dark matter is one of the first glimpses of possible physics beyond the Standard Model.  It is probable that dark matter consists of some stable or extremely long-lived particle left over from the hot early universe. For any given extension of the Standard Model containing stable or long-lived particles, the present-day relic density of such particles can prove quite constrainting.  The basic calculation of a relic density from thermal considerations is standard~\cite{KolbTurner}.  Here we review the steps in such a calculation under the explicit assumption that the lightest neutralino constitutes the cold dark matter particle.

The initial number
density of the neutralino is determined because the
particle is assumed to have been in thermal equilibrium.  When the particle begins in
thermal equilibrium with its surroundings, interactions that
create neutralinos usually happen as frequently as reverse
interactions which destroy neutralinos.  Once the temperature
drops below $T\simeq m_{\chi}$, most particles no longer have
sufficient energy to create neutralinos.  Now neutralinos can only
annihilate, and these annihilations occur until about the time
when the Hubble expansion parameter becomes larger than the
annihilation rate, $H\geq \Gamma_{\rm ann}$.  When expansion
dwarfs annihilation, neutralinos are being separated apart from
each other too quickly to maintain equilibrium.  This
happens at the freeze-out temperature, usually at $T_{F}\simeq
m_{\chi}/20$ for cold dark matter.

In most neutralino relic density calculations, the only interaction cross
sections that need to be calculated are annihilations of the type
$\chi \chi \rightarrow X$ where $\chi$ is the lightest neutralino and $X$ is any final state involving
only Standard Model particles.  However, there are scenarios in
which other particles in the thermal bath have important effects
on the evolution of the neutralino relic density.   Such a
particle annihilates with the neutralino into Standard Model particles and
is called a coannihilator~\cite{GrSe91}. To serve as an effective coannihilator, the particle must have direct
interactions with the neutralino and must be nearly
degenerate in mass.  Such degeneracy happens in the MSSM, for
instance, with possible coannihilators being the lightest
stau~\cite{Stau}, the lightest stop~\cite{Stop}, the
second-to-lightest neutralino or the lightest chargino
~\cite{Birkedal-Hansen:2001is,Birkedal-Hansen:2002am}.  When this
degeneracy occurs, the neutralino and all relevant coannihilators form a
coupled system.  In this section, we will denote particles
belonging to that coupled system by $\chi_{i}$.  Now all
interactions involving particles in this coupled system come into
play, including $\chi_{i} \chi_{j}\rightarrow X$, $\chi_{i} X
\rightarrow \chi_{j} Y$, and $\chi_{i} \rightarrow \chi_{j} X$.
Here both $X$ and $Y$ denote states including Standard Model
particles.  Decays once again enter the calculation because the
coannihilators are generally not stable and eventually decay into
the lightest neutralino.

For the case without coannihilations, evolution of the relic particle
number density, $n$,  happens in accordance with the single species
Boltzmann equation
\begin{eqnarray}
\frac{dn}{dt} = - 3 H n -\langle \sigma v \rangle
\left[n^2-\(n^{\rm eq}\)^2\right] ,
\end{eqnarray}

\noindent
where $n^{\mathrm{eq}}$ is the equilibrium number density, $H$ is the Hubble parameter at time $t$, and $\langle \sigma v \rangle$ is the thermally averaged annihilation cross section.
The number density is modified by Hubble expansion and
by direct and inverse annihilations  of the relic particle.  The relic particle is assumed to be stable, so relic
decay is neglected.  In the above expression, we have also assumed
T invariance to relate annihilation and inverse annihilation
processes.

In the presence of coannihilators, the Boltzmann equation becomes
more complicated but can be simplified using the stability
properties of the relic particle and the coannihilators (using
$n=\sum_{i=1}^{N} n_{i}$).  Application of these simplifications leads to
\begin{eqnarray}
\frac{d n}{dt}=&-&3 H n -\sum_{i,j=1}^{N} \langle\sigma_{ij}
v_{ij}\rangle \left(n_{i} n_{j} - n_{i}^{\rm eq} n_{j}^{\rm
eq}\right) .
\end{eqnarray}

To a very good approximation, one can use the usual single species
Boltzmann equation for the case of coannihilations if the
following replacement is made for the thermally averaged cross
section:
\begin{eqnarray}
\langle \sigma v\rangle = \sum_{i,j} \langle \sigma_{ij} v_{ij}
\rangle \frac{n_{i}^{\rm eq}}{n^{\rm eq}}\frac{n_{j}^{\rm
eq}}{n^{\rm eq}} .
\end{eqnarray}

For scenarios involving coannihilations, the expression for the
thermally averaged cross section begins as a six dimensional
integral, seven dimensional including the integration over the
final state angle.  This integral has been conveniently put into
the form of a one-dimensional integral over the total squared
center of mass energy~\cite{EdGo97}.  Computer codes exist which
numerically perform the center of mass momentum integration and
the final state angular integration.  Some codes include only a
subset of all possible coannihilation
channels~\cite{JuKaGr96,DarkSUSY}, while the program {\tt
micrOMEGAs} includes all relevant coannihilation channels.
Analytical expressions for the cross sections after integration over final
state angle also exist in the
literature~\cite{analytical}.  In practice, the Boltzmann equation is generally not given a full numerical treatment since accurate approximate solutions exist~\cite{KolbTurner}.

Once the effective thermally averaged cross section has been
derived as a function of temperature, it is used to iteratively
determine the freeze-out temperature.  In practice, the
dimensionless inverse freeze-out temperature, $x_{F} =
m_{\chi}/T_{F}$, is calculated using
\begin{equation}
x_{F} = \ln \left(\frac{0.038\, g\, m_{\PL} \,m_{\chi} \,\langle
\sigma v \rangle }{\sqrt{g_{*} \,x_{F}}}\right) .
\end{equation}
Here $m_{\PL}$ is the Planck mass, $g$ is the total number of
degrees of freedom of the $\chi$ particle (spin, color, etc.),
$g_{*}$ is the total number of effective relativistic degrees of
freedom at freeze-out, and the thermally averaged cross section is
evaluated at the freeze-out temperature.  For most cold dark matter
candidates, $x_{F}\simeq 20$.

When the freeze-out temperature has been determined, the total
(co)annihilation depletion of the neutralino number density can be
calculated by integrating the thermally averaged cross section
from freeze-out to the present temperature (essentially $T = 0$).
The thermally averaged cross section appears in the formula for
the relic density in the form of the integral $J\left(x_{F}\right)
= \int_{x_{F}}^{\infty} \langle\sigma v \rangle \, x^{-2} dx$:
\begin{equation}
\Omega_{\chi} {\rm h}^2 = 40
\sqrt{\frac{\pi}{5}}\frac{h^2}{H_{0}^2}\frac{s_{0}}{m_{Pl}^{3}}
\frac{1}{\left(g_{*S}/g_{*}^{1/2}\right) \, J\left(x_{F}\right)}.
\end{equation}
Here $g_{*S}$ is the number of effective relativistic degrees of freedom contributing to the entropy of the universe and $h$ is the reduced Hubble parameter, defined by $H_0 = 100\, h\, {\rm \, km \, sec^{-1} \,
Mpc^{-1}}$.  The above expression is commonly given in the less explicit form:
\begin{equation}
\Omega_{\chi} {\rm h}^2 = \frac{1.07 \times 10^9 \,
GeV^{-1}}{g_{*}^{1/2} \, m_{\PL} \, J\left(x_{F}\right)}.
\end{equation}
This is the expression that one compares with an experimental determination of the dark matter abundance.

Recent evidence suggests~\cite{Pr00} that $\Omega_{\chi} \sim 0.3$
with ${\rm h}^2 \sim 0.5$. We will take as a conservative favored
region,
\begin{equation}
0.1<\Omega_{\chi}{\rm h}^2<0.3. \label{limomega}
\end{equation}
\noindent The lower bound comes from the requirement that
$\chi^0_1$ should at least form galactic dark matter, and the
upper bound is a very conservative interpretation of the lower
bound on the age of the Universe. Recently, WMAP~\cite{WMAP} has elegantly confirmed
the composition of the Universe to be 73\% dark energy and 27\% matter.
WMAP determines a total matter density $\Omega_M {\rm h}^2 =
0.135^{+0.008}_{-0.009}$ and a total baryon density $\Omega_B {\rm
h}^2 = 0.0224 \pm 0.0009$, from which one can extract a 2$\sigma$
range for the density of cold dark matter: $\Omega_{\rm CDM} {\rm
h}^2 = 0.1126^{+0.0161}_{-0.0181}.$  We do not use this new result in our analysis.  The effect of the WMAP measurement has been studied in the case of neutralino dark matter from mSUGRA~\cite{Baer:2003yh} and rSUGRA~\cite{Birkedal-Hansen:2003gy} and also in the case of scalar dark matter from 'little Higgs' theories~\cite{Birkedal-Hansen:2003mp}.  Let us stress that the
requirement of~(\ref{limomega}) should not be treated as a
constraint, but rather as an indication of the region preferred by
cosmological considerations. Theoretical
assumptions made to extract the a present relic density of
neutralino LSP need not hold.  In fact, the missing non-baryonic
matter in the universe may not consist of relic neutralinos at
all.

\subsubsection{The $ b \rightarrow s \gamma$ constraint}

Another observable where the supersymmetric contribution can be
important and measurable is the flavor changing decay $ b
\rightarrow s \gamma$~\cite{Bertolini}. In the Standard Model,
this process is mediated by virtual isospin $+1/2$ quarks and
$W$-bosons. In supersymmetric theories, the spectrum allows new
contributions involving loops of charginos and squarks or top
quarks and charged Higgs bosons. As these two contributions appear
at the same order of perturbation theory, the measurement of the
inclusive decay $B \rightarrow X_s \gamma$ is a powerful tool in
the search for physics beyond the Standard Model. For our
analysis, we will use the results given by the CLEO and BELLE
collaborations~\cite{Cleo}.

The Particle Data Group~\cite{PDG} summarizes these
results and gives the current limit as:
\begin{equation}
\mathrm{BR} (b \rightarrow s \gamma)=(3.37 \pm 0.37 \pm 0.34 \pm
{0.24}_{-0.16}^{+0.35} \pm 0.38) \times 10^{-4} ,
\label{bsgamlimPDB}
\end{equation}
where the three first errors represent respectively the
statistical errors, the systematic error, and the estimated error
on the model describing the behavior of the quarks in the B-meson
decay. The fourth uncertainty is the error made by an
extrapolation to the entire energy range for the photon (cut at
2.1 GeV for the experiment). The last error value is an estimate
of the theoretical uncertainties. To be as conservative as
possible, one could add linearly all the uncertainties
of~(\ref{bsgamlimPDB}). But we will adopt the procedure taken in
the recent benchmark study of Battaglia et al.~\cite{bench} and
choose to impose the constraint
\begin{equation}
2.33 \times 10^{-4} < \mathrm{BR} (b \rightarrow s \gamma) < 4.15
\times 10^{-4} .
\label{bsgamlimus} \end{equation}
Let us note that we perform these calculations under the
assumptions of minimal flavor violation.

\subsubsection{The muon anomalous magnetic moment}

The relation between the spin $s$ of the muon and its magnetic
moment $\mu$ is given classically by $\mu=g_{\mu} \frac{e
\hbar}{2m_{\mu}c}s$ where $g_{\mu}=2(1+a_{\mu})$.
Following~\cite{MaWe03} we introduce the parameter $\delta_{\mu}$
to quantify the difference between theoretical and experimental
determinations of $a_{\mu}$:
\begin{equation}
\delta_{\mu} \equiv (a_{\mu} - 11~659~000 \times 10^{-10})\times
10^{10} .
\end{equation}

Recently, the Brookhaven collaboration has given a new measurement
of the anomalous magnetic moment of the muon~\cite{Brown}
\begin{equation}
\frac{(g^{\rm exp}_{\mu}-2)}{2} = a_{\mu}^{\mathrm{exp}}
=(11~659~202 \pm 14 \pm 6) \times 10^{-10},
\label{muonexp} \end{equation}
where the first error is statistical uncertainty and the
second is systematic uncertainty. From this the current
experimental determination of the parameter $\delta_{\mu}$ is
$\delta_{\mu}^{\mathrm{exp}}= 203 \pm 8$.

To understand the implications of this result for supersymmetry
one needs to know the Standard Model contribution. Unfortunately,
the computation of $\delta_{\mu}^{\SM}$ in the Standard Model is
complicated by our poor understanding of the hadronic
contributions to certain vacuum polarization diagrams. The
theoretical result depends on whether one uses the $e^+ e^-$
annihilation cross section or $\tau$ decay data to estimate these
contributions~\cite{Davier}:
\begin{equation}
\delta_{\mu}^{\mathrm{exp}}-\delta_{\mu}^{\mathrm{SM}}=(33.7 \pm
11.2) \times 10^{-10} ~~~~~[e^+e^-]
\end{equation}
\begin{equation}
\delta_{\mu}^{\mathrm{exp}}-\delta_{\mu}^{\mathrm{SM}}=(9.4 \pm
10.5) \times 10^{-10} ~~~~~[\tau\mathrm{-decay}] ,
\end{equation}
corresponding to a discrepancy between the Standard Model
prediction and experiment of respectively 3.0 and 0.9 standard
deviations. In our discussion we will be less conservative than
the authors of~\cite{MaWe03} and consider a 2 standard deviation
region about the anomalous moment of the muon based on the $\tau$
decay analysis:
\begin{equation}
 -11.6~<~\delta_{\mu}^{\mathrm{new \, physics}} =
 \delta_{\mu}^{\mathrm{exp}}-\delta_{\mu}^{\mathrm{SM}}~<~30.4 ~~~~~
 [2~\sigma]
\label{g-2}
\end{equation}

The contribution of SUSY particles to the anomalous moment of the
muon mainly comes from neutralino-smuon and chargino-sneutrino
loop-induced processes. As the chargino-slepton-lepton couplings
have the same form as the chargino-squark-quark couplings, the
parameter space restriction imposed by the constraint (\ref{g-2})
will be similar in many respects to the one given by the
$b\rightarrow s\gamma$ process. Thus, the SUSY contribution to
$(g_{\mu}-2)$ is large for large values of $\tan\beta$ and small
values of the soft-breaking masses. It is interesting to note that
the sign of the SUSY contribution is equal to the sign of the
$\mu$ parameter. With our conventions this implies that positive
$\mu$ is favored, as in the $b\rightarrow s\gamma$
constraint~\cite{Bertolini}.

Let us briefly summarize the above-mentioned constraints that are
to be applied below.  First, we demand that electroweak symmetry
be correctly broken, as defined by deriving a positive value for
$\bar{\mu}^2$ from~(\ref{radmuterm}).  It should be noted that we
use parameters of the Standard Model as defined
in~(\ref{SUSPECTferm}) and~(\ref{SUSPECTalpha}).  Next, we demand
that the mass of the lightest scalar Higgs boson be greater than
or equal to $113.5 \GeV$.  Additionally, we require that the mass
of the lightest chargino be greater than or equal to $103.5 \GeV$.
We also require that the lightest slepton mass be larger than $88
\GeV$ and the lightest squark mass be greater than $97 \GeV$, but
these two scalar fermion mass limits exclude no regions of
parameter space in the plots resulting from our analysis.  We also
indicate regions of parameter space that result in the right
thermal abundance of relic neutralinos: $0.1 \leq \Omega_{\chi}
h^2 \leq 0.3$.  While we technically do not exclude any regions
solely for failing to produce the cosmologically preferred relic
density, we do require that the lightest neutralino be the LSP. In
terms of the flavor changing decay $b \rightarrow s \gamma$, we
take a conservative approach and require the branching ratio to
obey $2.33 \times 10^{-4} < \mathrm{BR} (b \rightarrow s \gamma) <
4.15 \times 10^{-4}$, given in equation~\ref{bsgamlimus}. Finally,
we follow the recent $\tau$ decay analysis of the muon anomalous
magnetic moment and require $-11.6 < \delta_{\mu}^{\mathrm{new \,
physics}} < 30.4$, given above in equation~\ref{g-2}.

\section{Results and analysis}
\label{sec:phenom}

We are now in a
position to investigate the way in which low energy observations
can distinguish different regions of the weakly-coupled heterotic
string parameter space. We first
recall the standard analysis in the context of
the minimal supergravity (mSUGRA)
parameter space. While the unified mSUGRA paradigm is unlikely to
find a manifestation in realistic string-based models, it
nonetheless provides a useful benchmark for calibrating results
from one study to the next.

\subsection{Revisiting mSUGRA}
\label{sec:mSUGRA}

The minimal supergravity paradigm is based on the assumption that
the soft supersymmetry breaking Lagrangian is determined by only
five parameters. These include a universal gaugino mass $M_{1/2}$,
a universal scalar mass $M_0$ and a universal trilinear soft
parameter $A_0$, as well as the sign of the $\mu$ parameter and
$\tan\beta$. A quick survey of the soft terms presented in
Section~\ref{sec:space} indicates that there is no point in
the parameter space of the string models that we consider
that generates such a universal outcome. A
unified scenario is possible only in the event that the tree-level
dilaton domination case arises~\cite{dildom}, though no explicit
realization of this outcome exists in a complete model. Even if
such a scenario were realized, the soft terms would be constrained
to a particular point of the mSUGRA parameter space such that
\begin{equation}
-A_0 = M_{1/2} = \sqrt{3} M_0 .
\label{dildom} \end{equation}

\begin{figure}[t]
    \begin{center}
\centerline{
       \epsfig{file=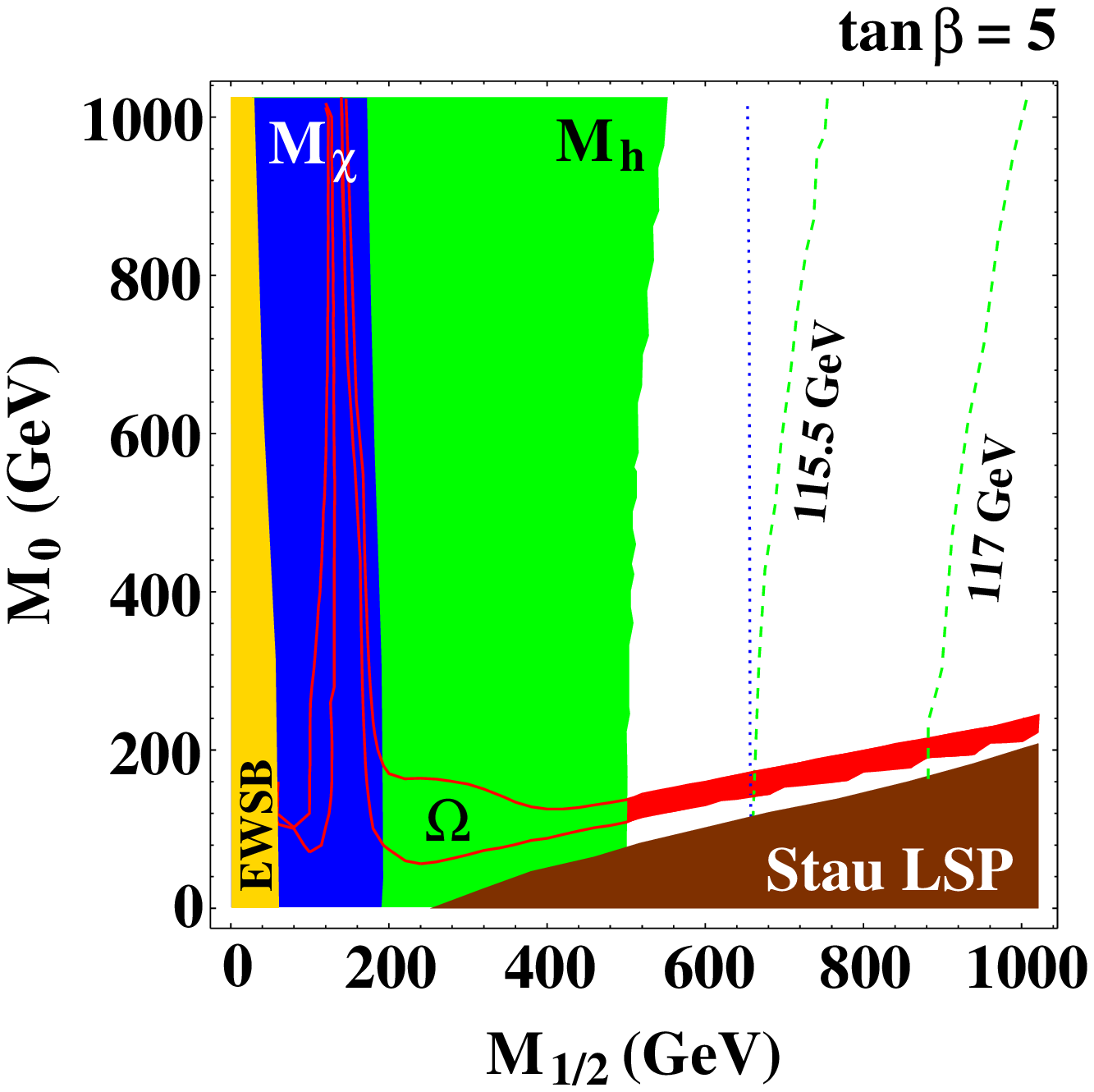,width=0.45\textwidth}
       \epsfig{file=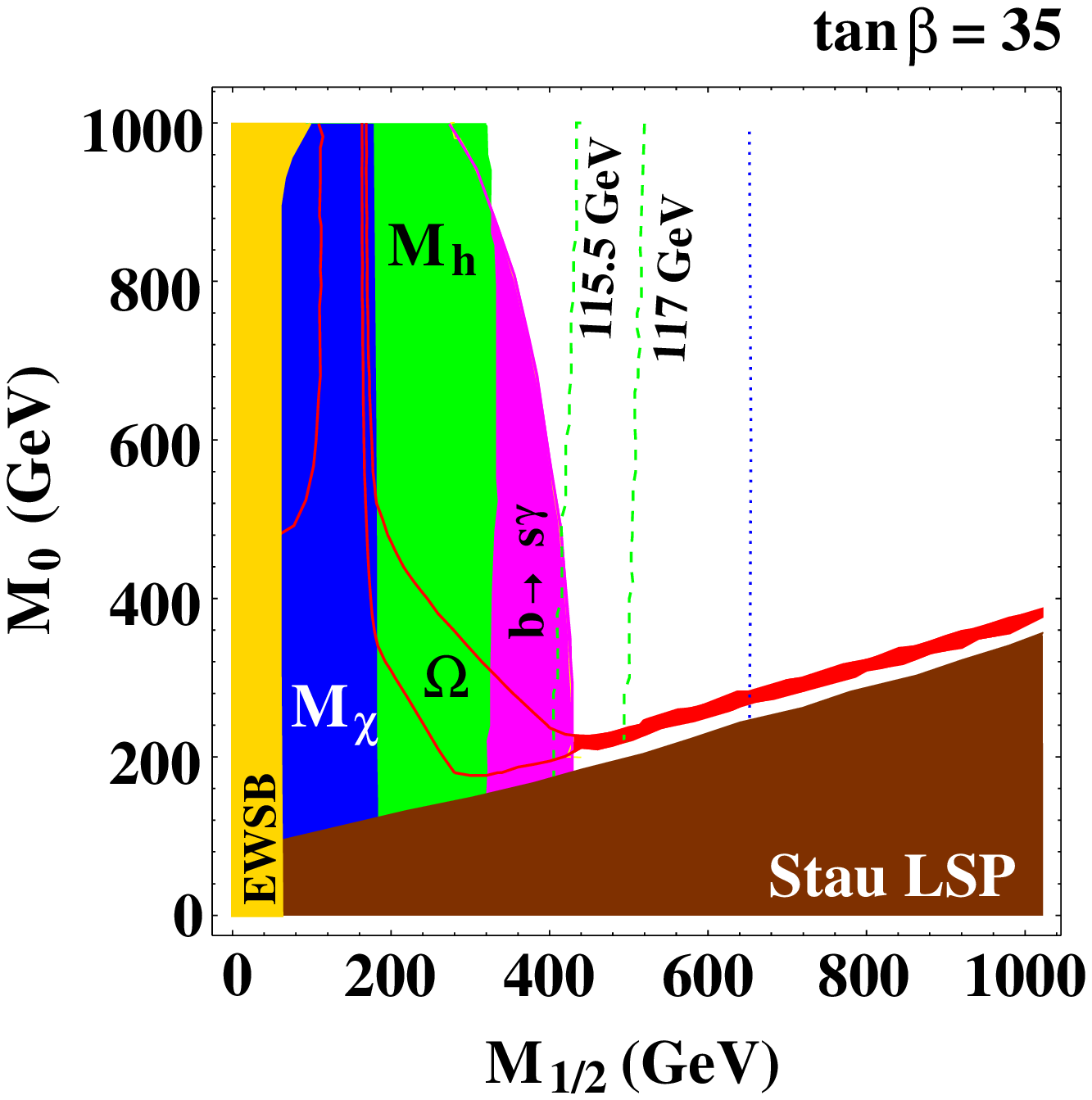,width=0.45\textwidth}}
          \caption{{\footnotesize {\bf Constraints on the mSUGRA parameter space for
          $\tan\beta=5$ (left) and $\tan\beta=35$ (right)}. Constraints on the
          $(M_{1/2},\; M_0)$ mSUGRA plane are given for $\mu > 0$ and $A_0=0$. The
          red contours and shaded region represent the $0.1 <\Omega_{\chi} {\rm h}^2
          <0.3$ preferred region. A small region at very low gaugino masses marked
          ``EWSB'' is ruled out by improper EWSB. Contours of constant Higgs mass of
          $m_h = 115.5 \GeV$ and 117 GeV (the dashed lines from left to right) as well
          as a contour of chargino mass $m_{\chi^{\pm}_{1}} = 500 \GeV$ (dotted line)
          are given. For a description of
          the experimental constraints applied, see Section~\ref{sec:analysis}.}}
        \label{fig:msugra5}
    \end{center}
\end{figure}

\begin{figure}[t]
    \begin{center}
\centerline{
       \epsfig{file=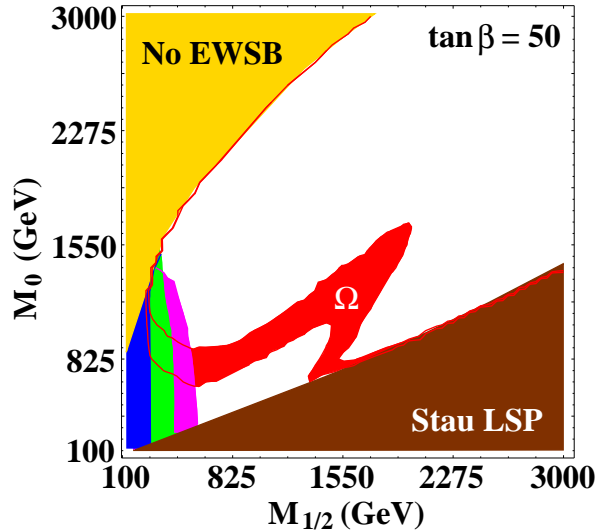,width=0.5\textwidth}}
          \caption{{\footnotesize {\bf Constraints on the mSUGRA parameter space for
          $\tan\beta=50$}. Constraints on the $(M_{1/2},\; M_0)$ mSUGRA plane are given
          for $\mu > 0$ and $A_0=0$. Note the change in scale relative to
          Figure~\ref{fig:msugra5}. The red region represents the $0.1 <
          \Omega_{\chi}{\rm h}^2 <0.3$ preferred region. The three
          shaded regions in the lower left are excluded by (from left to
          right) the chargino mass limit, the lightest Higgs boson
          mass limit and the $b \to s \gamma$ rate.}}
        \label{fig:msugra50}
    \end{center}
\end{figure}

Despite this lack of theoretical motivation, the extreme
simplicity of the mSUGRA approach makes it an attractive scenario
for phenomenological study~\cite{mSUGRA}. We will here just
present the main outcomes of our analysis. There are still large
areas of parameter space in the $(M_{1/2},\; M_0)$ plane which are
allowed, particularly for large $\tan\beta$ where the Higgs mass bound
is less constraining. The constraint from the process $b
\rightarrow s \gamma$ can be the dominant constraint for low
gaugino masses. Nevertheless, it is never very severe for positive
$\mu$, even for high $\tan\beta$ (note the change in scale in
Figure~\ref{fig:msugra50} relative to Figure~\ref{fig:msugra5}).
The situation would change in the case of large trilinear A-terms
because of the possibility of light third generation squarks~\cite{Stop}.
However, in this case charge and/or color breaking (CCB) problems
often arise.

In both plots of Figure~\ref{fig:msugra5} the exclusion regions
for the chargino and Higgs masses are given. These exclusion
regions implement the experimental limits of $103.5$ and $113.5
\GeV$, respectively, that were discussed in
Section~\ref{subsect:elect}. We also provide contours of constant
Higgs mass of $m_h = 115.5 \GeV$ and 117 GeV (the dashed lines
from left to right) as well as a contour of chargino mass
$m_{\chi^{\pm}_{1}} = 500 \GeV$ (dotted line). For small and
moderate values of $\tan\beta$ the requirement that the density of
the lightest neutralino accounts for the dark matter density is
very constraining and sensitive to the precise value of
$\tan\beta$. For moderate and low values of this parameter, one
usually finds too much relic density in the mSUGRA model. For
example the ``bulk'' region at low scalar and gaugino masses is
excluded by the Higgs mass limit. Only three regions survive: a
narrow band along the $\wtd{\tau}$-LSP exclusion zone, a large
Higgs pole region at extremely high $\tan\beta$, and the ``focus
point''~\cite{focus} region at high values of $M_{0}$ (the thin
strip along the ``no EWSB'' region in Figure~\ref{fig:msugra50}).
The region next to the $\wtd{\tau}$-LSP exclusion zone provides
for efficient $\wtd{\tau}_1 \wtd{\chi}^0_1$ coannihilation,
considerably reducing the neutralino density.  For high values of
$\tan \beta$, interesting areas exist where $\Omega_{\chi}{\rm
h}^2$ becomes smaller due to near-resonant s-channel annihilation
through the heavy Higgs states ($A$ or $H$). Here $m_A$ and $m_H$
become smaller and their couplings to the $b$ quark and the $\tau$
lepton increase.  In this region $\wtd{\chi}^0_1 \wtd{\chi}^0_1
\rightarrow b \overline{b}, \tau^+ \tau^-$ dominates and causes
significant depletion of the relic density. This can clearly be
seen in the plot for $\tan\beta$ = 50. It should be noted that
these results are strongly dependent on the treatment of the
radiative correction for the bottom mass and the Higgs masses. As
stated earlier, we have taken $M_b =4.62\GeV$. Small changes in
these values have dramatic consequences on the density. At large
$M_0$ and small $M_{1/2}$, there is another region (the "focus
point" region) where the lightest neutralino as well as the
lightest chargino acquire relatively small mass and large Higgsino
components. The coupling to the bosons $W$ and $Z$ is large enough
to make the annihilation process into $WW$ or $ZZ$ efficient.

Our results are in agreement with the previous studies of
mSUGRA~\cite{mSUGRA}. The only discrepancies appear in the high
$\tan\beta$ regimes, mostly due to the different treatment of the
$b-$quark mass and associated Yukawa coupling, and the Higgs
sector masses.  These differences can be understood in the light
of the work done by Allanach et al.~\cite{AlKrPo03}. Indeed, the
treatment of the Yukawa coupling of the $b-$quark ($\lambda_b$) in
{\tt SuSpect} results in a higher value for $\lambda_b$ than from
the other publicly available codes (a 4 to 8 percent effect). This
discrepancy affects the Higgs masses ($m_A$ in particular) via the
RGEs which are dominated by $\lambda_b$ for high $\tan\beta$. For
instance, for $\tan\beta=50$, we can easily find a 20 to 50 GeV
difference for $m_A$, explaining differences in the parameter
space allowed by the relic density in the $A-$pole sector. In the
remainder of our paper we will restrict ourselves to $\tan\beta
\leq 35$. In this range, the discrepancies between codes are
smaller than 3 percent. In terms of relic densities, the
differences between results using the codes {\tt
DarkSusy}~\cite{DarkSUSY} and {\tt micrOMEGAs}~\cite{Micromegas}
come from both the $b$-mass treatment and the selection of
included coannihilation channels: previous versions of {\tt
DarkSusy} did not include any coannihilation channels with
sleptons and scalars while {\tt micrOMEGAs} does.\footnote{The
upcoming version of {\tt DarkSusy} will include these
coannihilation channels.}

To summarize, in the case of the mSUGRA model, even if large areas
of the $(M_{1/2},M_0)$ parameter space are still allowed by
present accelerator data, combining the $b \rightarrow s \gamma$
limits, the neutralino relic density constraints and the Higgs
mass limit selects the $\wtd{\tau}$ coannihilation region, the
``focus point'' region, or extremely high values of $\tan\beta$.
These three regions represent the extreme boundaries of mSUGRA.
Thus, it is natural go beyond mSUGRA to see how these constraints
affect orbifiold models.

\subsection{Moduli domination: the anomaly-mediated limit}
\label{sec:AMSB}

The phenomenology of the original (minimal) AMSB model has been
extensively studied (for some of the earlier work,
see~\cite{AMSBref}). However the phenomenology of the string-based
PV-AMSB model, defined by the soft terms given in~(\ref{amsb}),
has yet to be thoroughly explored. We view this anomaly-dominated
regime as a particular limit of string models where supersymmetry
is broken by the F-term of some compactification modulus. Note
that this anomaly-mediated limit is one in which all tree level
soft supersymmetry breaking terms vanish. We thus choose to begin
our study of string-based heterotic orbifold phenomenology at one
loop with this particularly simple regime.

\begin{figure}[tb]
    \begin{center}
\centerline{
       \epsfig{file=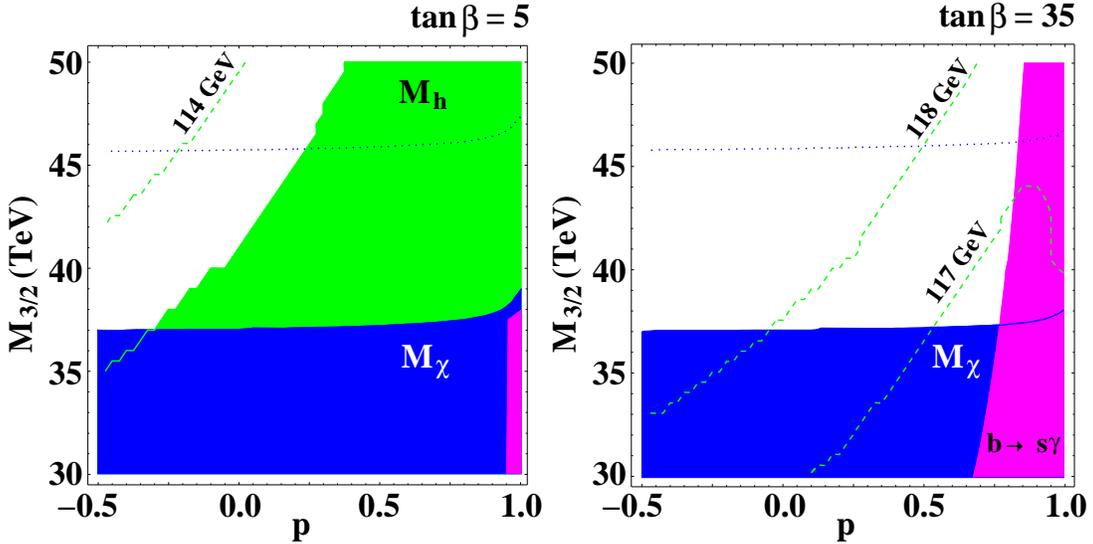,width=0.9\textwidth}}
          \caption{{\footnotesize {\bf Constraints on the PV-AMSB parameter space for
          $\tan\beta=5$ (left) and $\tan\beta=35$ (right)}. Constraints on the
          $(M_{3/2}, p)$ plane are given for $\mu > 0$. The regularization weight $p$
          can be as large as $p=1$, at which point scalar masses vanish at the one
          loop level (the standard AMSB limit). The areas marked $M_{\chi}$ and $M_h$
          are ruled out by the chargino mass bound and the Higgs mass bound,
          respectively. The area marked $b \to s \gamma$ gives
          ${\rm Br}(b \to s \gamma) \geq 4.15 \times 10^{-4}$ (a small region where
          this bound is relevant occurs near $p=1$ for $\tan\beta=5$). The horizontal
          (dotted) contour is a constant chargino mass of $m_{\chi^{\pm}_{1}} = 130 \GeV$.}}
        \label{fig:AMSBpM}
    \end{center}
\end{figure}

Like the minimal AMSB scenario, the PV-AMSB model has two free
parameters, apart from $\tan\beta$ and the sign of the $\mu$ term.
One parameter is the overall scale, given by the gravitino mass
$M_{3/2}$. The other parameter is the regularization weight $p$
which determines the size of the scalar mass terms relative to the
gaugino masses and A-terms.  In this sense the weight $p$ plays a
role comparable to the bulk $M_0$ postulated in the minimal AMSB
scenario but it is not an {\em ad hoc} parameter and it appears
elsewhere in the spectrum. Note however that the gaugino masses
are independent of this weight. Thus achieving a sufficiently
heavy chargino implies a lower bound on the absolute scale as
displayed in Figure~\ref{fig:AMSBpM}.

If we look at the branching ratio $b \rightarrow s \gamma$ we
observe than it becomes important in two situations: small values
of the gravitino masses and/or large values of $p$. As previously
mentioned, small gravitino masses give small values of
$m_{\chi^{\pm}_1}$, which leads to a big contribution of the
diagram involving $\chi^{\pm} \tilde q$ in the loop for $b
\rightarrow s \gamma$.  Having a large value for $p$ also leads to
small squark masses and thus a large contribution from the squark
exchange diagrams.  Moreover, the same limit $p\to 1$ also leads
to light $\tilde \chi^0_{3,4}$, $\tilde \chi^{\pm}_2$, which can
now also contribute substantially to $b \rightarrow s \gamma$
(because of the small $\mu$ parameter).

Concerning the Higgs mass, our findings agree with intuition
gained from mSUGRA. One needs significant loop contributions from
relatively heavy squarks to satisfy the LEP constraint, especially
at low $\tan\beta$. Thus regions where squarks are light are ruled
out. This includes the region with $p \to 1$ and smaller gravitino
mass scales. In Figure~\ref{fig:AMSBpM} we show the region
excluded by our Higgs mass constraint in the left plot, as well as
a contour of $m_h = 114 \GeV$ for comparison. In the right plot
for $\tan\beta=35$ there is no constraint in this gravitino mass
range arising from the Higgs search limit and we provide contours
of $m_h = 117 \GeV$ and 118 GeV.

In previous AMSB studies it was found that the neutralino thermal
relic density is generically too small to explain the amount of
dark matter~\cite{AMSBdm}. Due to the low ratio of $M_{2}/M_{1}$
in the AMSB scenarios, the wino content of the LSP is quite high.
Additionally, coannihilation between the LSP and the lightest
chargino is also very efficient.  Both of these effects combine to
make the thermal relic density of LSP negligible.  Throughout the
parameter space of Figure~\ref{fig:AMSBpM} the thermal relic
density is around $\Omega_{\chi} {\rm h}^2 \sim 10^{-4}$. Thus the
anomaly-mediated character of the gaugino sector in this model
necessitates a non-thermal production mechanism for neutralino
LSPs, or another candidate for the cold dark matter must be
postulated.

\begin{figure}[tb]
    \begin{center}
\centerline{
        \epsfig{file=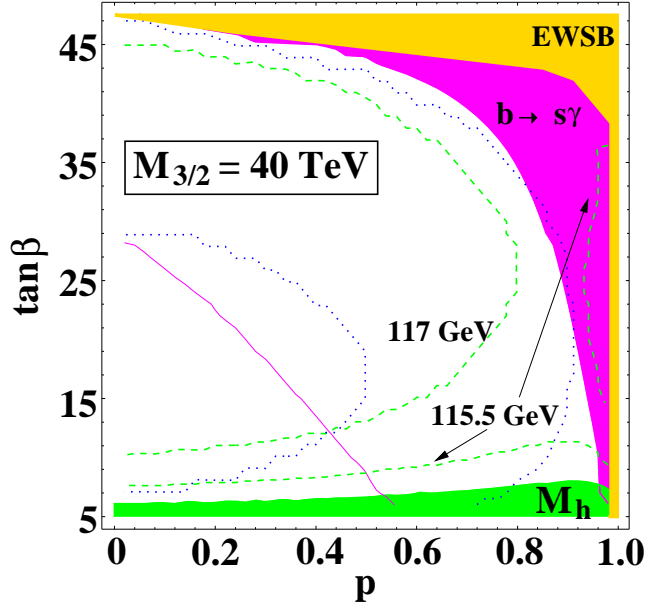,width=0.55\textwidth}}
          \caption{{\footnotesize {\bf Constraints on the PV-AMSB parameter space for
          $M_{3/2} = 40 \TeV$}. Constraints on the
          $(p, \; \tan\beta)$ plane are given for $\mu > 0$. Regions excluded by
          EWSB failure (top),
          excessive rate for $b \to s \gamma$ (right) and the Higgs mass limit (bottom)
          are shaded. We have labeled contours of $m_h = 115.5 \GeV$ and 117 GeV (dashed
          lines), as well as unlabeled contours of $m_{\chi^{\pm}_1} = 112.5 \GeV$ and
          113.5 GeV (dotted lines). The solid contour in the lower left is a contour
          of ${\rm BR}(b \to s \gamma) = 3.53 \times 10 ^{\-4}$.}}
        \label{fig:AMSBpTAN}
    \end{center}
\end{figure}

We sum up these constraints in Figure~\ref{fig:AMSBpTAN} where we
have shown the allowed region in the $(p, \; \tan\beta)$ plane for
a gravitino mass of $M_{3/2} = 40 \TeV$.  We clearly see that high
$\tan \beta$ area is excluded by $b \rightarrow s\gamma$, and the
low $\tan\beta$ area is disfavored because of the Higgs mass
limit. In every case, $p$ cannot be too large (must be less than
$\sim 0.85$). This excludes the ``minimal'' AMSB scenario ($p=1$)
up to the two-loop corrections to soft terms which have not been
fully calculated in these string-based models. Most of the
interior of the parameter space in Figure~\ref{fig:AMSBpTAN} has
sufficiently heavy Higgs masses, as shown by the dashed contours
of $m_h = 115.5 \GeV$ and 117~GeV. The supersymmetric
contributions to the $b\to s \gamma$ rate in the interior are
small as well. The solid contour is ${\rm BR} (b \to s \gamma) =
3.53 \times 10^{-4}$. The chargino masses are very light
throughout and we have provided contours of $m_{\chi^{\pm}_1} =
112.5 \GeV$ (right-most dotted line) and 113.5 GeV (left-most
dotted line). We note that much of this parameter space predicts a
relatively small supersymmetric contribution to the muon anomalous
magnetic moment due to relatively heavy scalar masses ({\em i.e.}
$\delta_{\mu}^{\SUSY} \sim 0$). The upper shaded region is ruled
out by improper electroweak symmetry breaking ($\mu^2 < 0$).

In summary, the Pauli--Villars regularization weight $p$ generates
a whole class of anomaly-dominated models at the one-loop level.
While this new degree of freedom solves the tachyonic slepton
problem, the minuscule relic density still does not allow the
neutralino to be a thermal dark matter candidate. Additionally,
the constraints coming from accelerator physics exclude high
values of $p$ and low values of $M_{3/2}$ and predict small
contributions to the muon anomalous magnetic moment.

\subsection{The general moduli dominated case}
\label{sec:modplots}

We will now relax the assumption that the various K\"ahler
moduli are stabilized at self-dual points, while maintaining the
assumption that the modular weights of all matter fields are equal
and given by $n_{i} = -1$. When the vacuum value $\lang \re\;t
\rang$ is not fixed at $\lang \re\;t \rang = 1$ we must use the
soft terms in~(\ref{modsoft}). Note that the value of the
Green-Schwarz coefficient $\delta_{\GS}$ becomes relevant for the
determination of gaugino masses in this case. For the time being
we will take $\delta_{\GS}= 0$ in order to keep the parameter
space to manageable size.

For the study of the moduli dominated case we have looked at the
behavior of the model as a function of the vacuum expectation value ($vev$) of the real part
of the (universal) modulus field $t=T|_{\theta=0}$, for different
values of $p$ and $\tan\beta$. To be more precise, we have taken
two values of $p$, 0 and 0.95, and two values of $\tan\beta$, 5 and 35.
We have chosen to use the value $p=0.95$ because at this value the
gaugino masses and scalar masses have roughly the same magnitude.
It is also very near the pure AMSB limit. We first stress the
common features of the moduli-dominated case, and extract the
corresponding phenomenological consequences, before looking at
specific points in the parameter space.

\begin{figure}[tb]
    \begin{center}
\centerline{
       \epsfig{file=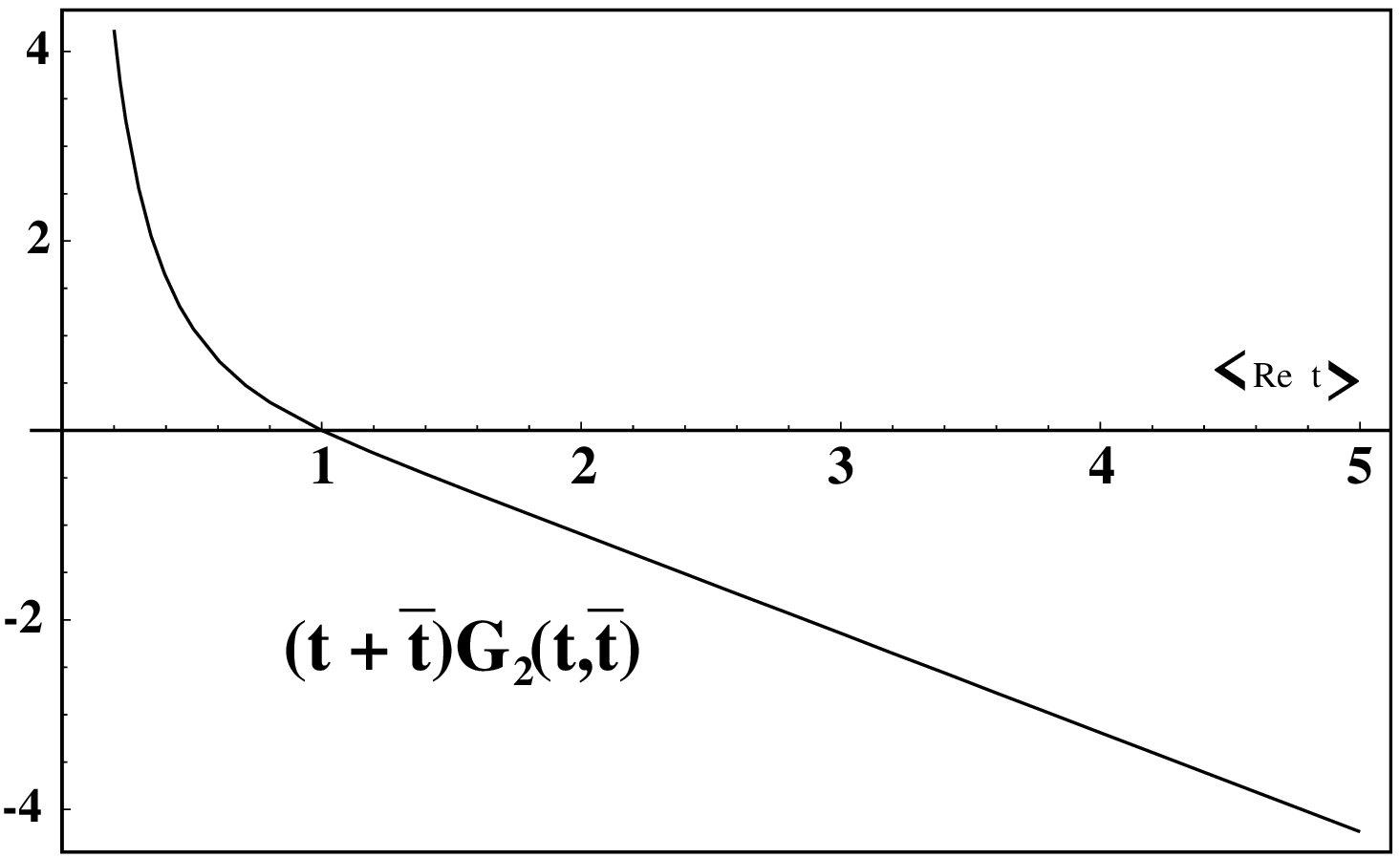,width=0.5\textwidth}}
          \caption{{\footnotesize {\bf Behavior of the combination $(t+\bar{t}) \Eisen$
          as a function of $\re \; t$}. }}
        \label{fig:modfunc}
    \end{center}
\end{figure}

We choose to investigate the region $0.1 \leq \lang \re \; t \rang
\leq 5$. Since the vacuum value of this modulus is related to the
radius of the compact dimensions, negative values have no physical
meaning. Let us note that the combination $(t+\bar{t}) \Eisen$,
which appears in the gaugino mass when the equation of
motion~(\ref{cos}) is substituted into~(\ref{modsoft}), is
invariant (up to a sign) under the duality transformation $t \to
1/t$ as shown in Figure~{\ref{fig:modfunc}. However, since this
term competes with the anomaly-mediated contribution to gaugino
masses at the one loop level, gaugino masses will not display this
duality symmetry. Any value $\lang \re \; t \rang \neq 1$
represents a spontaneous breaking of modular invariance and the
phenomenology of the theory for $\lang \re \; t \rang < 1$ will be
different from the case $\lang \re \; t \rang > 1$.

Looking again at the gaugino masses in~(\ref{modsoft}) it is clear
that when $\delta_{\GS} = 0$ the gaugino masses are proportional
to beta-function coefficients. Thus we expect the phenomenology of
this scenario to be very similar to that of the PV-AMSB scenario
of the previous section. For example, we continue to have
\begin{equation}
\left.\frac{M_1}{M_2}\right|_{\rm GUT} =\frac{g_1(\mu_{\rm
GUT})}{g_2(\mu_{\rm GUT})}\frac{b_1}{b_2} ~ \sim ~
\frac{b_1}{b_2} = \frac{33}{5} = 6.6
\end{equation}
independent of the value of $\lang \re \; t \rang$ or $p$. At the
low scale this implies $M_2 \ll M_1$ and the lightest neutralino
$\chi^0_1$ and the lightest chargino $\chi^{\pm}_1$ are in a
nearly complete Wino state. This means that for all
values of $p$ and $\lang \re \; t \rang$ we expect $m_{\chi^0_1}
\simeq m_{\chi^{\pm}_1} \simeq M_2$, and the relic density of the
LSP neutralino is almost entirely depleted by (co)annihilation
channels involving  $\chi^0_1$ and $\chi^{\pm}_1$.

Unlike the case of Section~\ref{sec:AMSB}, however, there is now a
value of $\lang \re \; t \rang$ such that the contribution from
the Eisenstein function exactly counterbalances the contribution
from the superconformal anomaly giving us vanishing gaugino mass
soft breaking terms $M_a$ to this order. For the choice of phase
conventions in~(\ref{Faux}) and~(\ref{Maux}) this occurs
when\footnote{In~\cite{BiGaNe01} the same behavior was noted. In
that reference, however, the opposite sign convention
on~(\ref{Faux}) was used, leading to vanishing gaugino masses when
$(t+\bar{t})\Eisen = -1$ which occurs at the dual point $\re \; t
\simeq 2$.}
\begin{equation}
(t+\bar{t})\Eisen = 1 \quad \rightarrow \quad \zeta(t)=0 \quad
\rightarrow \quad \re \; t = 0.523 .
\label{zetazero} \end{equation}
Not surprisingly, then, we find that much of the parameter space
in the vacinity of  $\lang \re \; t \rang = 0.5$ is ruled out by
the chargino mass bounds for any reasonable choice of scale
$M_{3/2}$ in Figures~\ref{fig:modp0} and \ref{fig:modp095}.

\begin{figure}[tb]
    \begin{center}
\centerline{
       \epsfig{file=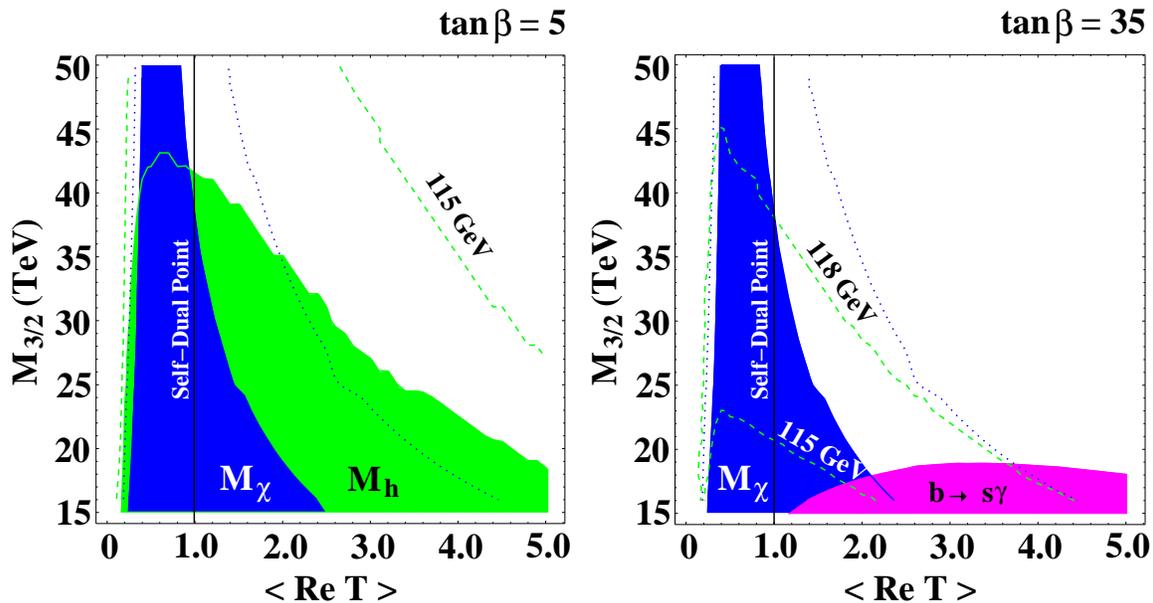,width=0.95\textwidth}}
          \caption{{\footnotesize {\bf Constraints on the moduli-dominated
          parameter space for $\tan\beta=5$ (left) and $\tan\beta=35$ (right) with
          $p = 0$}. Constraints on the ($M_{3/2},\; \lang \re \; t \rang$) plane
          are given for $\delta_{\GS} = 0$ and $\mu > 0$. The region in $\lang \re \;
          t \rang$ shown here is $0.2 \leq \lang \re \; t \rang \leq
          1$. The areas marked $M_{\chi}$ and $M_h$
          are ruled out by the chargino mass bound and the Higgs mass bound,
          respectively. Chargino masses of $m_{\chi^{\pm}_1} = 200 \GeV$ (dotted line)
          are provided in both plots, while Higgs contours of $m_h = 115
          \GeV$ and 118 GeV (dashed lines) are given where applicable.}}
        \label{fig:modp0}
    \end{center}
\end{figure}

In the extreme cases where $\lang \re \; t \rang \to 0.1$ and
$\lang \re \; t \rang \to 5$ the absolute value of the combination
$(t+\bar{t})\Eisen$ becomes large and dominates over the anomaly
contribution in the gaugino masses. Here gauginos will typically
be similar in size to the scalar masses at the high scale,
particularly as the value of $p$ is increased towards its limiting
value $p=1$ (Figure~\ref{fig:modp095}). The chargino mass bound is
easily satisfied here and the increased contribution from $M_3$ in
the RG evolution of the scalar masses makes the Higgs mass
constraint easier to satisfy, particularly for large $\tan\beta$.
In Figure~\ref{fig:modp0} we show the Higgs and chargino exclusion
regions in the $\tan\beta=5$ panel and the contours $m_h = 115
\GeV$ (dashed line) and $m_{\chi^{\pm}_1} = 200 \GeV$ (dotted
line). The Higgs mass is not constraining at $\tan\beta=35$ and we
provide contours of $m_h = 115 \GeV$ and 118 GeV (dashed lines)
and again $m_{\chi^{\pm}_1} = 200 \GeV$ (dotted line).

\begin{figure}[tb]
    \begin{center}
\centerline{
       \epsfig{file=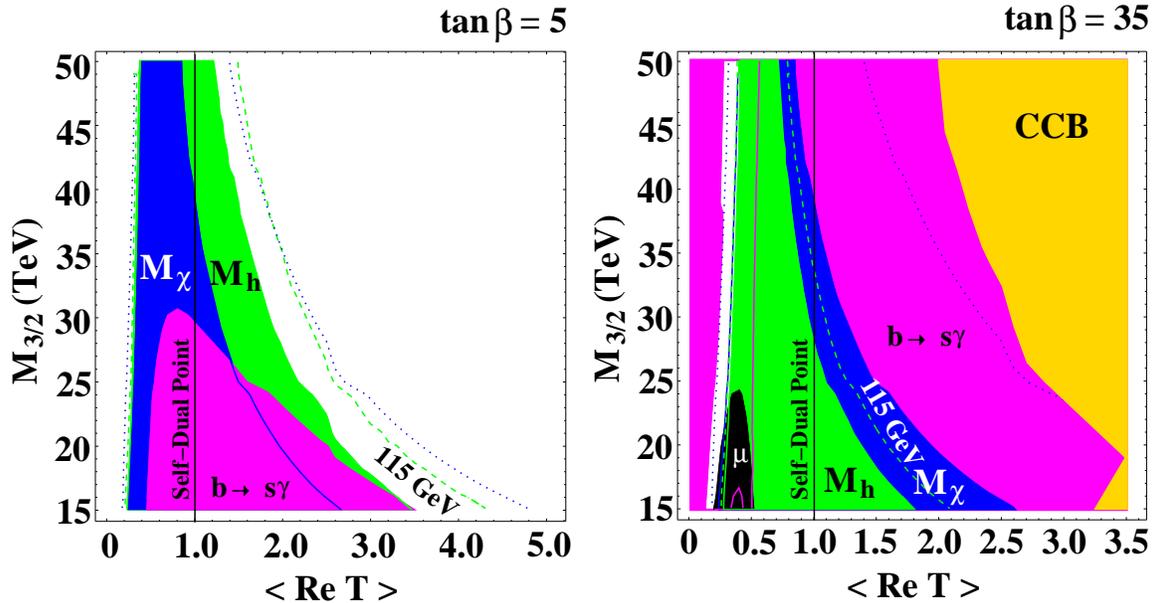,width=0.95\textwidth}}
          \caption{{\footnotesize {\bf Constraints on the moduli-dominated
          parameter space for $\tan\beta=5$ (left) and $\tan\beta=35$ (right) with
          $p = 0.95$}. Constraints on the ($M_{3/2},\; \lang \re \; t \rang$) plane
          are given for $\delta_{\GS} = 0$ and $\mu > 0$. The shaded region in the
          upper right of the $\tan\beta=35$ plot is excluded by CCB constraints, while
          the snall shaded triangle in the lower left is excluded by improper EWSB.
          Chargino mass contours of $m_{\chi^{\pm}_1} = 200 \GeV$ (dotted line)
          and Higgs mass contours of $m_h = 115 \GeV$ (labeled dashed line) are given
          in both plots.}}
        \label{fig:modp095}
    \end{center}
\end{figure}

While the mass contours and exclusion regions in
Figure~\ref{fig:modp0} show the underlying modular symmetry $t \to
1/t$ in a very broad sense, it is clear that these contours do not
obey this symmetry in the strict sense. We have indicated the
self-dual point $\lang \re \; t \rang = 1$ by a heavy vertical
line. The exclusion regions do not show the expected duality
symmetry about this point. As already mentioned above, this is due
in large part to the competition between contributions to the
gaugino masses from the moduli sector and contributions from the
superconformal anomaly. The relative sign between these
contributions depends on which side of the self-dual point the
(overall) modulus is stabilized. Having chosen the conventions
of~(\ref{Faux}),~(\ref{gaugtree}) and~(\ref{Apot}) this implies a
zero in the gaugino mass formula~(\ref{Maloop}) on a definite side
of the self-dual point (with the conventions here, the smaller
value side). The conventions can be changed to make this zero
occur on the other side, but only at the expense of changing the
convention for the sign of the $\mu$ parameter. As we keep this
sign always fixed at positive values, the {\em relative} sign
between the gaugino masses -- in particular, the gluino mass $M_3$
-- and the $\mu$ term is the key variable in distinguishing the
two sides of the self-dual point. This is reflected in the fact
that at $\tan\beta=35$ in Figure~\ref{fig:modp0} the $b \to s
\gamma$ constraint applies to only one side of the figure. Other
observables, such as the muon anomalous magnetic moment, will be
sensitive to this relative sign as well, making any spontaneous
breaking of modular invariance a true observable, at least in
principle.\footnote{For a discussion of these points, in the
context of distinguishing anomaly mediation from gauge mediation,
see the work of G.~Kribs in~\cite{AMSBref}.}

As we approach the minimal AMSB limit of $p \to 1$ the relevant
constraints change as scalar masses diminish. This is displayed in
Figure~\ref{fig:modp095}. The Higgs mass limit is important over
much of the parameter space, requiring gravitino mass scales well
in excess of $50 \TeV$ for $\lang \re \; t \rang \simeq 1$ in
order to compensate for the rapidly diminishing stop masses. Note
that Figures~\ref{fig:modp0} and~\ref{fig:modp095} correctly
reproduce our previous results in the limit of $\lang \re \; t
\rang \to 1$. The light squarks in the $p \simeq 1$ regime tend to
contribute too much to the $b \rightarrow s \gamma$ process -- a
process that provided no such constraint in this region of $\lang
\re \; t \rang$ in Figure~\ref{fig:modp0} due to the relatively
large scalar masses in those cases.

Nearly the entire region of the parameter space for $\tan\beta=35$
shown in Figure~\ref{fig:modp095} is ruled out by this constraint.
The remaining slender allowed region near $\lang \re \; t \rang
\simeq 0.25$ allowed by the experimental constraints is the result
of a relatively light charged Higgs that helps to cancel the
contributions from the light chargino. Here the scalars and the
gauginos are light -- in particular, the gluino has a vanishing
soft mass at the high scale near $\lang \re \; t \rang \simeq 0.3$
when $\delta_{\GS} = 0$ (see Figure~\ref{fig:gluinomass} below).
This leads to a very small $\mu$ term value in this narrow range,
and consequently a small charged Higgs mass. For example, as we
move from $\lang \re \; t \rang \simeq 0.1$ to $\lang \re \; t
\rang = 0.3$ along the line $M_{3/2} = 15 \TeV$ in the right panel
of Figure~\ref{fig:modp0} the value of $\mu$ at the electroweak
scale drops from 1700 to 1100 GeV due to the diminishing gluino
mass. But in the right panel of Figure~\ref{fig:modp095} $\mu$
goes from 1700 to 380 GeV over the same range, leading to a
charged Higgs mass of approximately the same value. Scalar masses
are around 200 GeV at this point.

In the right panel of Figure~\ref{fig:modp095} with $p \to 1$
(light squarks) and large $\tan\beta$ (large couplings in the
Higgs sector) there also appears a zone excluded by too large of a
supersymmetric contribution to the anomalous magnetic moment of
the muon (a black region labeled ``$\mu$'') due to the presence of
very light squarks and gauginos. This region is already excluded,
however, by multiple observables such as the $b \to s \gamma$
rate, the chargino mass and the Higgs mass bound. The shaded
region in the upper right of the $\tan\beta=35$ plot in
Figure~\ref{fig:modp095} is disfavored by the presence of CCB
minima deeper than the preferred electroweak vacuum. The presence
of such a region is due to the very large trilinear couplings for
these values of $\lang \re \; t \rang$ which are larger than the
scalar masses when $p \simeq 1$.

In conclusion we see that in this simple moduli-dominated scenario
the phenomenology is similar to that of anomaly mediation. The
closer one approaches the minimal AMSB limit of $p=1$ the more
problematic the scenario becomes, particularly at large
$\tan\beta$. Other points in the PV-AMSB class have some areas of
phenomenological viability, though the chargino and Higgs mass
bounds can be quite constraining, particularly near the self-dual
point $\lang \re \; t \rang =1$. All of these cases will be unable
to explain the cold dark matter content of the universe, at least
in terms of thermal relic neutralinos.

Our study of the PV-AMSB scenario of Section~\ref{sec:AMSB} had
one effective parameter, the regularization weight $p$, apart from
the overall scale given by the gravitino mass $M_{3/2}$. This
regularization weight controls the size of the scalar mass
relative to the fundamental scale $M_{3/2}$ while the gaugino
masses were fixed by their beta-function coefficients. In the
present section the parameter space has been expanded to allow for
$\lang \re \; t \rang \neq 1$, allowing the gaugino masses to be
varied relative to the scalars and gravitino mass $M_{3/2}$ in an
independent way. Small, or even vanishing, gaugino masses are now
possible, as is the possibility of relatively large gaugino masses
as one moves far from the self-dual point for the compactification
moduli.

Despite this new degree of freedom the parameter space is still
quite constrained. In particular, obtaining the correct relic
density for the neutralino LSP is impossible in this framework,
because of its extreme wino--like nature. But, as we will see in
the next section, the possibility of non-vanishing Green-Schwarz
counterterm can open a new region of parameter space completely in
accord with present experimental data.

\subsection{The influence of the Green-Schwarz counterterm}
\label{sec:gsplots}

In this section we wish to introduce the possibility that
$\delta_{\GS} \neq 0$, as would typically be the case in realistic
heterotic orbifold constructions. As we will see, this parameter
allows one to interpolate between regions with a phenomenology
similar to anomaly mediated models (specifically the PV-AMSB
scenario of equation~\ref{amsb}) and regions where the
phenomenology resembles that of the minimal supergravity
paradigm.

\begin{figure}[tb]
    \begin{center}
\centerline{
       \epsfig{file=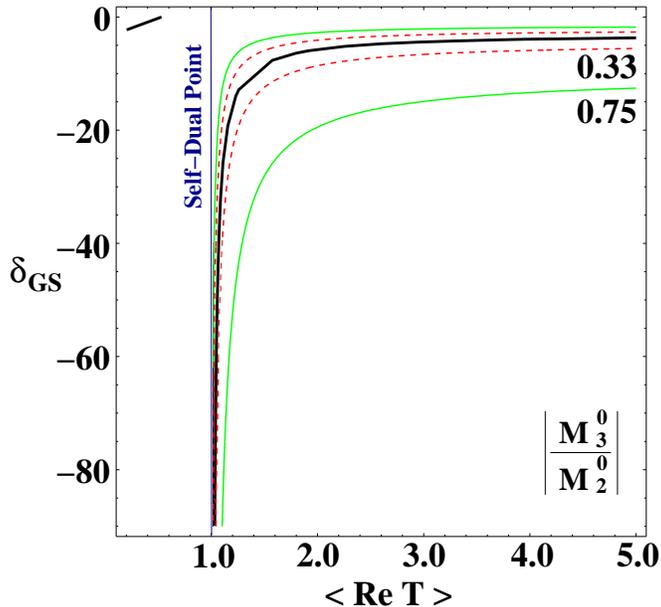,width=0.55\textwidth}}
          \caption{{\footnotesize {\bf Contours of relative running gaugino masses
          $M_3/M_2$ in the $(\lang \re \; t \rang ,\,\; \delta_{\GS})$ plane}. These soft
          masses are at the initial (GUT) scale. The heavy (dark) contour is the limit
          of vanishing gluino mass (there is another such contour in the upper left
          corner on the other side of the self-dual point). For $\lang \re \; t \rang\ >1$
          we also give contours of $|M_3/M_2| = 0.33$ (dashed) and 0.75 (solid).}}
        \label{fig:gluinomass}
    \end{center}
\end{figure}

In the conventions defined by the string threshold
correction~(\ref{floop}) the value of $\delta_{\GS}$ is a negative
integer between 0 and -90. For any given orbifold construction the
precise value of this coefficient can be worked out, and indeed
such an exercise has been performed for realistic $Z_3$ models
already~\cite{Gi01b}. In the figures that follow we will treat
this variable as a continuous parameter. At values of the K\"ahler
modulus far from its self-dual point, where the Eisenstein
function is not negligible, the term in~(\ref{modsoft})
proportional to $\delta_{\GS}$ can give rise to substantial
(universal) contributions to the gaugino masses. Thus points in
parameter space with large values of $\delta_{\GS}$ and values of
$\lang \re \; t \rang$ far from unity will likely have a pattern
of gaugino masses similar to mSUGRA -- and hence are more likely
to provide a viable dark matter candidate than cases with
$\delta_{\GS} \simeq 0$. Furthermore, once $|\delta_{\GS}| \sim
\sqrt{16\pi^2} \sim \order (10)$, the gaugino masses and scalar
masses will be of approximately the same size. This is very
different from the hierarchical situation of the previous
sections.

Because of the difference of sign between the beta functions $b_a$
of the Standard Model gauge groups, $M_3$ has a particular
behavior that differs from the other gaugino masses. This in turn
drives the phenomenology of the model, particularly through the
gluino's RG effects on the rest of the superpartner spectrum. With
a non-vanishing Green-Schwarz coefficient the various gaugino
masses will no longer have zeros at the same value of  $\re \; t
$. In particular, for certain combinations of $\lang \re \; t
\rang$ and $\delta_{\GS}$ it is possible for the gluino mass to
vanish at the boundary condition scale while the other gauginos
have non-vanishing masses. These combinations are shown in
Figure~\ref{fig:gluinomass}. For parameter choices near these
contours we might expect the phenomenologically unacceptable
result of a gluino LSP. Even when the gluino is not the LSP, its
small value relative to other gauginos may provide too little
radiative correction to squark masses and the Higgs mass bound may
be difficult to satisfy. On the other hand, a relatively light
gluino is likely to reduce the amount of fine-tuning required to
obtain $M_Z = 91.2 \GeV$~\cite{KaLyNeWa02}.

\begin{figure}[tb]
    \begin{center}
\centerline{
       \epsfig{file=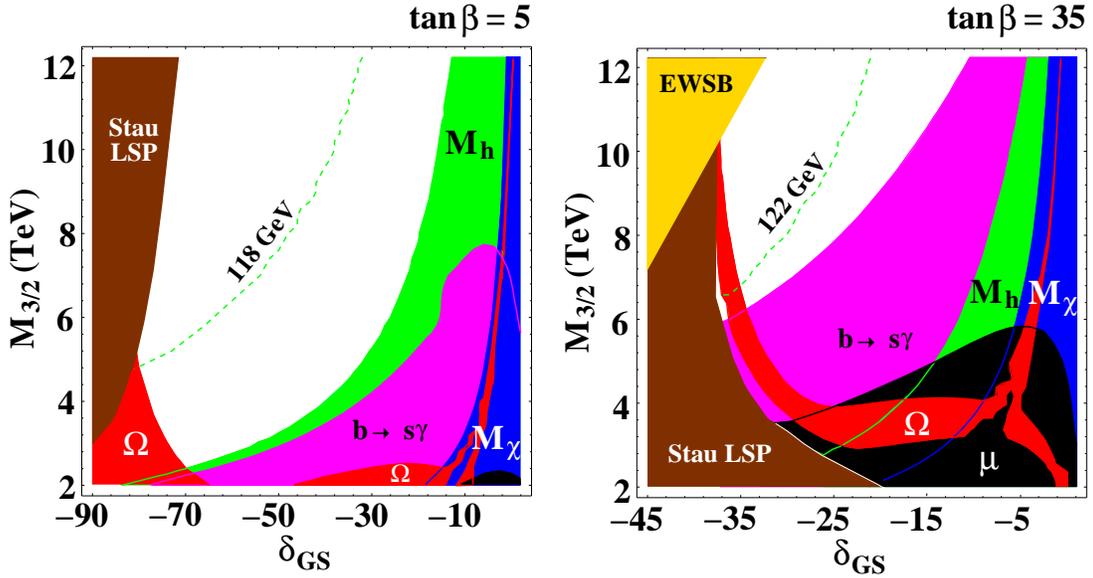,width=0.9\textwidth}}
          \caption{{\footnotesize {\bf Constraints on the moduli-dominated
          parameter space for $\tan\beta=5$ (left) and $\tan\beta=35$ (right) with
          $p = 0$ and $\lang \re \; t \rang = 0.5$}. Constraints on the
          ($M_{3/2},\; \delta_{\GS}$) plane are given for $\mu > 0$. Shaded regions
          to the left are ruled out by improper EWSB or a stau LSP. The dark region
          in the lower right of the $\tan\beta=35$ plot labeled ``$\mu$'' has too
          much supersymmetric contribution to $(g_{\mu} -2)$. The strip labeled ``
          $\Omega$'' has the cosmologically preferred neutralino relic density. Dashed
          contours of $m_h = 118 \GeV$ (left) and $m_h = 122 \GeV$ (right) are also
          given.}}
        \label{fig:modp0t05}
    \end{center}
\end{figure}

In Figures~\ref{fig:modp0t05}, \ref{fig:modp0t123}
and~\ref{fig:modp0t2} we look at this model in the
$(\delta_{\GS},\; M_{3/2})$ plane for $p=0$ and $\lang \re \; t
\rang = 0.5$, $\lang \re \; t \rang = 1.23$ and $\lang \re \; t
\rang = 2$, respectively. Beginning with the case $\lang \re \; t
\rang = 0.5$ in Figure~\ref{fig:modp0t05} we see that the limit as
$\delta_{\GS} \to 0$ reproduces the cases studied in the previous
section. In particular, such models require very large $M_{3/2}$
in these regions to satisfy the chargino and Higgs mass
constraints. This behavior is seen in all three of the  $\lang \re
\; t \rang$ presented here. A maximum value of $|\delta_{\GS}|$
can in general be obtained by requiring that lightest neutralino
be heavier than the lightest stau. For any values of $\tan\beta$
the maximum value of $|\delta_{\GS}|$ is determined by the
requirement of keeping the lightest neutralino as the LSP. The
quantity $|\delta^{\mathrm{max}}_{\mathrm{GS}}|$ is smaller for
higher $\tan\beta$ because the stau is in this case lighter.

In between these disallowed regions the increasing values of the
gaugino masses for fixed scalar mass, as the Green-Schwarz
coefficient increases in absolute value, results in smaller
contributions to the $b \rightarrow s \gamma$ branching ratio and
a smaller (in absolute value) anomalous magnetic moment of the
muon. In Figure~\ref{fig:modp0t05} we have chosen a value of
$\lang \re \; t \rang$ near the value where gaugino masses vanish
when $\delta_{\GS} \to 0$. As is evident from the figure, gaugino
masses are indeed falling as this limit is reached. Note that in
this case there is no point in the plane for which the gluino is
the LSP.

\begin{figure}[tb]
    \begin{center}
\centerline{
       \epsfig{file=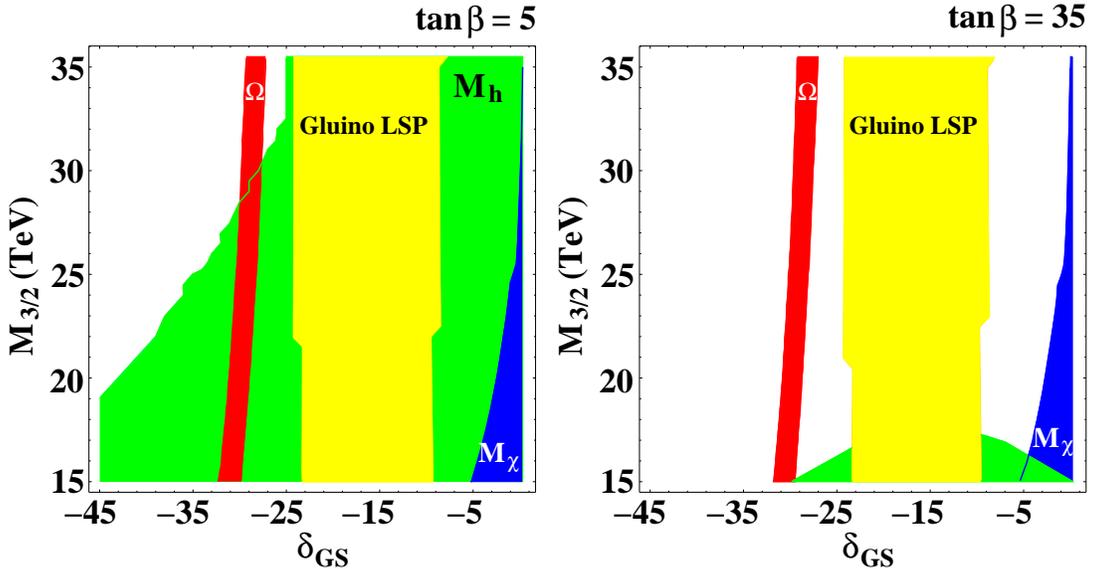,width=0.9\textwidth}}
          \caption{{\footnotesize {\bf Constraints on the moduli-dominated
          parameter space for $\tan\beta=5$ (left) and $\tan\beta=35$ (right) with
          $p = 0$ and $\lang \re \; t \rang = 1.23$}. Constraints on the
          ($M_{3/2},\; \delta_{\GS}$) plane are given for $\mu > 0$. The lightly
          shaded region in the center gives a gluino LSP. The thin strip with
          acceptable thermal neutralino relic density is labeled ``$\Omega$.'' Note the
          change in scale from Figure~\ref{fig:modp0t05}.}}
        \label{fig:modp0t123}
    \end{center}
\end{figure}

Concerning the relic density, it is negligible for small values of
the coefficient for the Green--Schwarz counterterm, as we have
seen previously: the neutralino is extremely Wino-like in nature
and is nearly degenerate with the lightest chargino. Almost all of
the relic neutralinos are depleted through coannihilation channels
with the charginos. When we increase $|\delta_{GS}|$, however, we
increase the gaugino masses through the increased importance of
the $G(t,\overline t)$ terms in~(\ref{modsoft}). This universal
contribution can come to compete with the (nonuniversal)
anomaly-mediated term, driving the gaugino mass terms $M_{i=1,2}$
far away from the previous ``degenerate'' situation. The gaugino
mass sector then begins to look similar to that of mSUGRA and a
more bino-like LSP can develop, leading to an increased relic
density. Just as in the mSUGRA case there continues to be a region
of stau coannihilation near the excluded stau-LSP area. And --
again, just as in mSUGRA -- much of the ``bulk'' area is ruled out
by the chargino and Higgs mass bounds, or the $b\to s \gamma$
constraint.

In Figure~\ref{fig:modp0t123} we investigate the same parameter
space for the case where $\lang \re \; t \rang = 1.23$. When
multiple gaugino condensates are utilized to stabilize the
dilaton, with the tree level K\"ahler potential given by
$K(S,\oline{S})=-\ln(S+\oline{S})$, then a modular invariant
treatment of the resulting non-perturbative potential for the
moduli fields will lead to the conclusion that the auxiliary field
for the dilaton $F^S$ must vanish in the vacuum. The remaining
potential for the K\"ahler moduli leads to their stabilization at
this value~\cite{FoIbLuQu90}. Specific points in the space shown
in Figure~\ref{fig:modp0t123} were singled out for a more detailed
study of their collider signatures in a recent set of
string-inspired ``benchmark'' models~\cite{ourbench}. Here a more
complete survey of the parameter space is possible.

\begin{figure}[tb]
    \begin{center}
\centerline{
       \epsfig{file=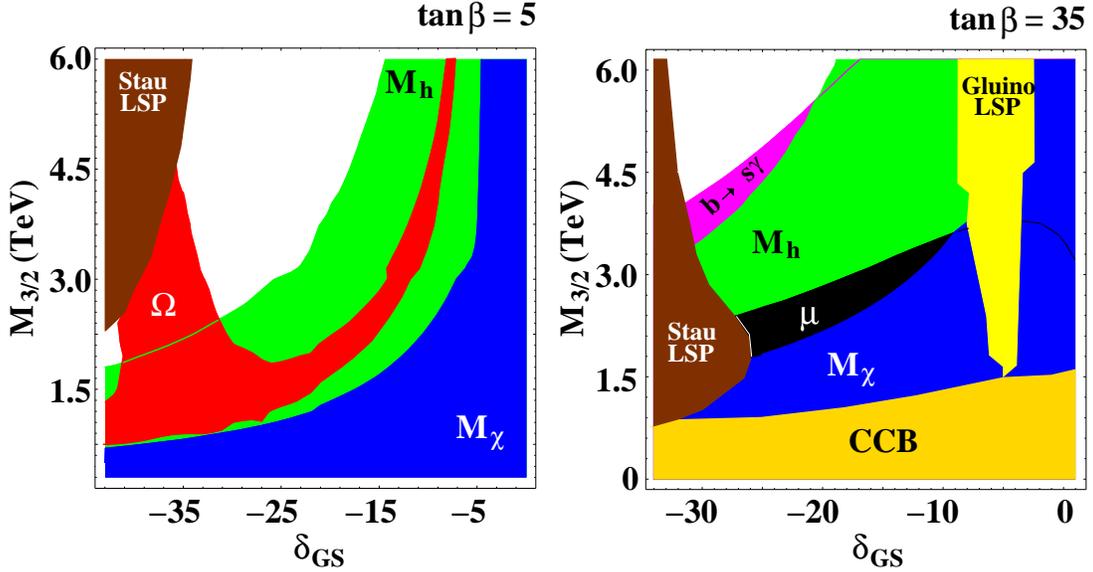,width=0.9\textwidth}}
          \caption{{\footnotesize {\bf Constraints on the moduli-dominated
          parameter space for $\tan\beta=5$ (left) and $\tan\beta=35$ (right) with
          $p = 0$ and $\lang \re \; t \rang = 2.0$}. Constraints on the
          ($M_{3/2},\; \delta_{\GS}$) plane are given for $\mu > 0$. The dark shaded
          regions on the left have a stau LSP. The $\tan\beta=35$ plot also has a
          region with a gluino LSP. For $\tan\beta=5$ the region labeled
          ``$\Omega$'' has the cosmologically preferred relic density of neutralinos.
          No such region exists for the higher $\tan\beta$ plot. In that case the
          exclusion contours are due to (from bottom right to upper left) CCB vacua,
          the chargino mass, too large SUSY contributions to $(g_{\mu}-2)$, the Higgs
          mass limit and too large a $b \to s \gamma$ rate. }}
        \label{fig:modp0t2}
    \end{center}
\end{figure}

In accordance with Figure~\ref{fig:gluinomass} we note a region
centered about $\delta_{\GS} = -15$ where the gluino is the LSP.
Interestingly, this is precisely the region of values favored by
semi-realistic $Z_3$ orbifold constructions~\cite{Gi01b}. To the
right, as the absolute value of the Green-Schwarz counterterm
diminishes, the chargino mass bound is eventually reached. The
value $\lang \re \; t \rang = 1.23$ is sufficiently close to the
self-dual value that the theory near $\delta_{\GS}=0$ is anomaly
mediated-like in character: very large values of the gravitino
mass are generally required to achieve a sufficiently heavy
lightest chargino and lightest Higgs (note the difference in scale
between Figures~\ref{fig:modp0t05} and~\ref{fig:modp0t123}). For
$\tan\beta=35$ the Higgs mass constraint is less important, being
only relevant for low gravitino masses near the gluino LSP region.
In addition, the LSP is predominantly wino-like on the right side
of the gluino LSP region, and the relic density of these
neutralinos is negligible.

As we move to larger absolute values of $\delta_{\GS}$ we
interpolate between this anomaly mediated region and an
mSUGRA-like region. For certain critical values of this parameter
the wino content of the LSP is sufficiently reduced to result in
the correct annihilation efficiency to produce the cosmologically
preferred relic abundance. This change in the character of the LSP
is the result of the increasing importance of moduli contributions
to the gaugino masses (the first term in the gaugino mass
parameter in equation~\ref{modsoft}) versus the anomaly mediated
contributions (the last term in that expression). The ability to
achieve the right relic density, independent of the scalar mass
values, is a general property of models with this sort of
non-degeneracy among gaugino mass
parameters~\cite{Birkedal-Hansen:2001is,Birkedal-Hansen:2002am}.

\begin{figure}[tb]
    \begin{center}
\centerline{
       \epsfig{file=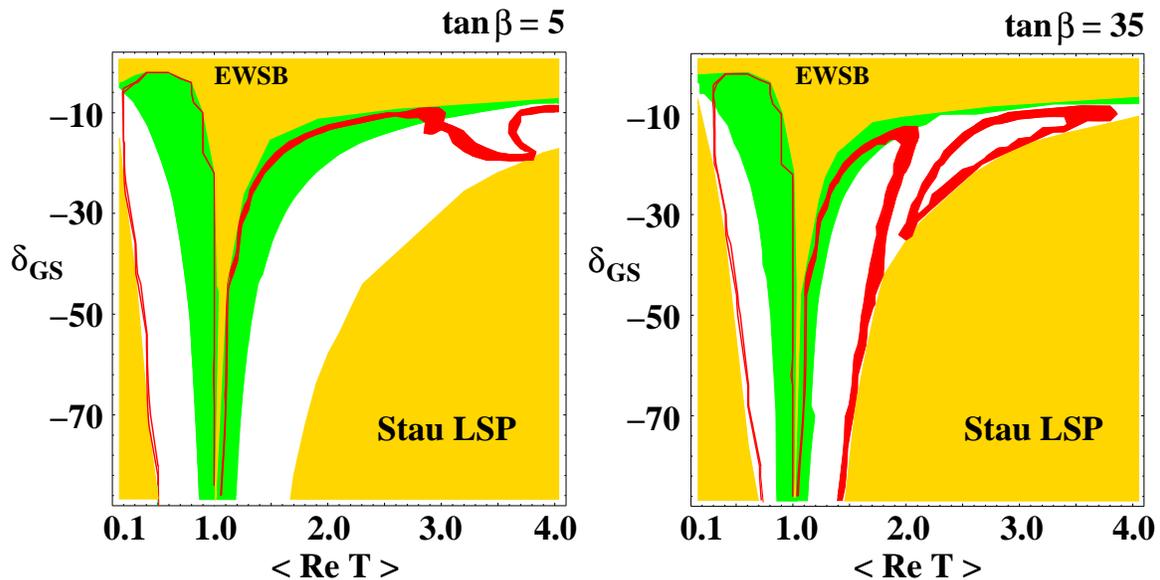,width=0.95\textwidth}}
          \caption{{\footnotesize {\bf Constraints on the moduli-dominated
          parameter space for $\tan\beta=5$ (left) and $\tan\beta=35$ (right) with
          $p = 0$ and $M_{3/2}=10 \TeV$}. Constraints on the
          ($\lang {\rm Re} t \rang,\; \delta_{\GS}$) plane are given for $\mu > 0$.
          The upper light shaded region has improper EWSB, while the lower light shaded
          regions on either side of $\lang \re \; t \rang =1$ have a stau LSP. All
          collider constraints have been combined in the dark shaded region to make
          the preferred relic density strip easier to distinguish.}}
        \label{fig:modp0tdGS}
    \end{center}
\end{figure}

We investigate the case $\lang \re \; t \rang = 2.0$ in
Figure~\ref{fig:modp0t2}. The behavior at small $\tan\beta$ is not
dissimilar from the nominally ``dual'' case of $\lang \re \; t
\rang = 0.5$ in Figure~\ref{fig:modp0t05}. The stau LSP constraint
now arrives at much smaller values of the Green-Schwarz
coefficient. Again, the region with the promising relic density of
neutralinos is largely constrained by the Higgs mass limit.

The parameter space for the higher $\tan\beta$ regime is even more
tightly constrained (at least at these relatively low values of
the gravitino mass). We again see a region of gluio LSP -- a
narrower region at lower absolute values of $\delta_{\GS}$
consistent with Figure~\ref{fig:gluinomass}. At the bottom of the
plot the values are ruled out by the presence of CCB vacua while
the region to the left has a stau LSP. The light charginos for
this value of the gravitino mass lead to large contributions to
the muon anomalous magnetic moment (the black region labeled
``$\mu$''). Finally, all but the uppermost corner of the space is
ruled out by the Higgs mass limit and the upper bound on ${\rm
BR}(b \to s \gamma)$.

We combine all of these observational and theoretical constraints
to explore the relation between $\delta_{\GS}$ and $\lang \re \; t
\rang$ in Figure~\ref{fig:modp0tdGS}. We plot the allowed
parameter space in the $(\delta_{\GS}, \; \lang \re \; t \rang)$
plane in a similar fashion to Figure~\ref{fig:gluinomass} for the
specific gravitino mass of $M_{3/2}=10 \TeV$. On either side of
the self-dual point at large negative values of $\delta_{\GS}$ the
stau is the LSP. Near these regions there is a strip of preferred
neutralino relic density. The area at small values of
$|\delta_{\GS}|$ and near $\lang \re \; t \rang=1$ is ruled out by
improper EWSB. Here again there is a small strip of preferred
relic density, though this is within the region ruled out by the
Higgs mass, chargino mass or $b\to s \gamma$ constraint. In the
interior there continues to be viable parameter space, including
points with acceptable relic neutralino densities.

\subsection{The generalized dilaton domination case}
\label{sec:dilplots}

If we now consider the other extreme case ($\theta=\pi/2$) the
general features are somewhat different. Referring to the soft
terms of equation~\ref{BGWsoft}, we can clearly see that the soft
scalar mass terms will be roughly dominated by the gravitino mass
scale (up to radiative corrections). We anticipate, therefore,
that the relevant gravitino mass scale will be approximately one
order of magnitude lower than the one necessary in the
moduli-dominated case because of the absence of the loop
suppression factors on the scalar masses. The gaugino mass soft
supersymmetry breaking terms will be determined by the dilaton
auxiliary field $vev$ $\lang F^S \rang$.

In the case where nonperturbative corrections to the dilaton
K\"ahler potential are imagined, the minimum of the combined
(modular invariant) dilaton/K\"ahler moduli potential now occurs
at $\lang F^S \rang \neq 0$ and compactification moduli stabilized
at self-dual points where $\lang F^T \rang =0$. As described in
Section~\ref{sec:dilaton}, the requirement of vanishing vacuum
energy naturally leads to a suppression of this auxiliary field
$vev$ relative to the auxiliary field of the supergravity
multiplet. Thus the two contributions to the gaugino mass in
equation~\ref{BGWsoft} each involve a loop suppression factor
(unlike the scalar masses) and the dilaton and conformal anomaly
contributions are comparable. Indeed, the beta--function
coefficients $b_a$ of the Standard Model gauge groups are of the
order of $10^{-2}$. Moreover, the $F^S$ term will be driven by
$b_+$, the largest beta--function coefficient among the condensing
gauge groups of the hidden sector, from~(\ref{FS})
and~(\ref{Ktrue}). In fact, if we look in more detail at the
expression for $F^S$ in equation~\ref{FS}, it is apparent that for
not too large values of $b_+$ we can consider that $F^S$ has a
linear evolution as a function of that parameter. Increasing $b_+$
has a direct consequence on the values of the gaugino breaking
terms, while increasing $M_{3/2}$ has direct consequences on the
general size of all soft breaking terms -- in particular soft
scalar masses. We can consider the effect of the parameter $b_+$
as a ``fine structure'' on top of the gross feature of a hierarchy
between scalars and gauginos in this regime.

\begin{figure}[tb]
    \begin{center}
\centerline{
       \epsfig{file=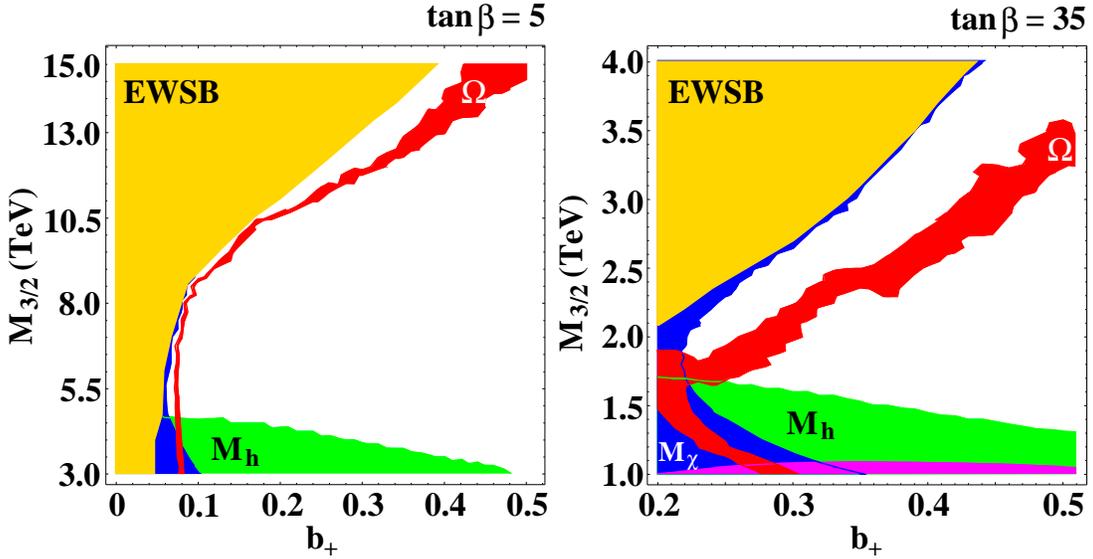,width=0.9\textwidth}}
          \caption{{\footnotesize {\bf Constraints on the dilaton-dominated
          parameter space for $\tan\beta=5$ (left) and $\tan\beta=35$ (right) with
           }. Constraints on the
          ($M_{3/2},\; b_+$) plane are given for $\mu > 0$.}}
        \label{fig:dilaton}
    \end{center}
\end{figure}

The general features of this situation are shown in
Figure~\ref{fig:dilaton}. When the suppression of the gaugino
masses is large from small values of $b_+$ (small $a_{\rm np}$),
the gaugino sector looks increasingly like that of
anomaly-mediation. This is true of any value of $\tan\beta$. As
this value is increased the wino content of the LSP diminishes and
we approach the overwhelmingly bino-like LSP of mSUGRA. However,
the value of $b_+$ cannot increase without limit. As we are
considering the weakly coupled $E_8 \times E_8$ heterotic string,
we cannot imagine a condensing group larger than the complete
$E_8$ of the hidden sector, for which $b_{E_8} = 90/16\pi^2 =
0.57$. It is more likely that the same Wilson-line mechanism that
breaks the observable sector gauge group to the product of
sub-groups at the compactification scale also breaks the hidden
sector group as well. Thus a more typical range for this parameter
might be $b_+ \leq 15/16\pi^2 = 0.095$, where this value
corresponds to a pure Yang-Mills sector of SU(5).

For this range of condensing group beta-function, only small
values of $\tan\beta$ are consistent with the stringent demands of
electroweak symmetry breaking. This fact was observed in this
model some time ago~\cite{GaNe00a}, but it is here displayed
explicitly in Figure~\ref{fig:dilaton}. The large regions marked
``EWSB'' in this figure have $\bar{\mu}^2 < 0$. This is a
manifestation of the ``focus point,'' or hyperbolic nature of this
model~\cite{focus}: the scalar masses are generally much larger
than gaugino masses throughout the parameter space. The light
gluinos imply that third generation squarks must have large masses
at the boundary condition scale to ensure a sufficiently heavy
Higgs mass. Thus $M_{3/2} \gappeq 4 \TeV$ is needed in the
$\tan\beta=5$ case and $M_{3/2} \gappeq 1.3 \TeV$ is needed in the
$\tan\beta=35$ case.

Note also that the very large (and nearly universal) scalar masses
will imply heavy sleptons at the electroweak scale. Therefore both
direct neutralino annihilation through slepton exchange, as well
as $\wtd{\tau}_1/\chi^0_1$ coannihilation, will be ineffective in
reducing the relic neutralino density in the early universe.
Nevertheless, acceptable relic densities are possible in this
model, without the need for resonant annihilation through a heavy
Higgs eigenstate. The cosmologically preferred region is the
shaded strip labeled ``$\Omega$'' in Figure~\ref{fig:dilaton}.
Between this strip and the region ruled out by improper EWSB the
relic density is below the preferred amount. In this region the
LSP has a sufficiently large wino content to annihilate
efficiently. Below the marked strip the LSP is bino-like and the
relic density is too large. We note that the region with the
correct relic density can continue indefinitely to large scalar
masses by maintaining the correct wino-content of the LSP.

\section*{Conclusion and Perspectives}
\label{sec:conclusion}

Even within the restricted context of the weakly-coupled heterotic
string compactified on an orbifold, the phenomenology of such
supergravity effective theories is far richer and varied than that
of the standard ``minimal'' supergravity approach. While our
current indirect knowledge on the nature of supersymmetry already
constrains these models significantly, there still exist large
regions of parameter space for all the cases studied here -- even
for low values of $\tan\beta$. We anticipate that future
measurements of, or limits on, superpartner masses and kinematic
distributions will constrain the parameters of these models
further. So too will future cosmological or astrophysical
measurements. In fact, these sorts of measurements form a useful
complementarity that will be crucial in unraveling the nature of
supersymmetry breaking and transmission in such models in the
manner begun here. Future studies of string-based models should
focus on extending this initial survey to true collider signatures
in both hadron and lepton machines, as well as computing event
rates for astrophysical processes.

Our analysis in this paper was somewhat cursory -- we consider the
constraints that arise principally from the chargino and Higgs
searches at LEP as well as the measurement of the branching ratio
for $b \to s \gamma$ in the minimal flavor violation scenario. We
have indicated, where applicable, the region that seems to be
favored by cosmological observations of dark matter relic
densities. Interpreting this observation as a constraint on a
model implies the assumption that the dark matter is the result of
thermal processes in the early universe and is composed entirely
of relic neutralino LSPs. In string models many other dark matter
candidates are known to exist: axions, superheavy exotic string
states, hidden sector gauge and matter condensates, etc., so care
should be exercised in applying the cosmological bounds. We have
therefore kept the range we show within the wider region $0.1 \leq
\Omega_{\chi} {\rm h}^2 \leq 0.3$ rather than the far more
restrictive result from the recent WMAP experiment. We have also
used only the most conservative possible interpretation of the
measurement of the muon anomalous magnetic moment as to not
prematurely prejudice one model versus another.

Despite the uncertainties that must be borne in mind in such an
analysis arising from the computational tools employed ({\em e.g.}
the choice of quark pole masses utilized, uncertainties in the
determination of $\bar{\mu}$ at large $\tan\beta$ etc.) some
general statements can certainly be made. Among these is the
observation that explicit models of moduli stabilization utilizing
field-theoretic methods tend to predict that at least some sector
of the theory obtains soft supersymmetry breaking terms only at
the one-loop level. This is especially true of the gaugino sector.
Furthermore, this suppression leads to the general requirement
that the gravitino be relatively heavy in these cases -- perhaps
alleviating the cosmological problems so often associated with
models that contain gravitinos.

Parameters that are related to the orbifold can have a large
impact on these broad features. This is a welcome result --
implying that experimental data can indeed probe the nature of the
underlying theory within a class of models. For example, in cases
where supersymmetry breaking is transmitted predominantly by the
moduli associated with compactification the relative sign of terms
proportional to their auxiliary field and that of the conformal
anomaly can in principle be measured. This parameter, in turn, is
related to the spontaneous breakdown of modular invariance that
may occur in such models.

The phenomenology, even in the simple initial approach taken here,
is neither that of minimal supergravity nor minimal anomaly
mediation but is often a hybrid of the two. This makes a detailed
study of collider signatures all the more urgent as search
strategies are often based on one or the other of these paradigms.
Given that the era of experimental supersymmetry may well be at
hand -- and that weakly coupled heterotic strings still represent
the best motivated string-based approach to understanding
supersymmetry and its breaking in the low-energy world -- the time
is right for this exciting undertaking.

\section*{Acknowledgements}
A. B.-H. gratefully acknowledges the hospitality of Laboratoire de
Physique Th\'eorique, Universit\'e Paris-Sud
were significant portions of this work were completed. Y.M. would like to thank
E Dudas and JB de Vivie for useful disussions. A. B.-H. is
supported in part by the DOE Contract DE-AC03-76SF00098 and in
part by the NSF grant PHY-00988-40.



\begin{thebibliography}{99}
\bibitem{GiLuMuRa98}
  {\rm G.~F.~Giudice, M.~Luty, H.~Murayama and R.~Rattazzi},
  {\it JHEP} {\bf 9812} {\rm (1998) 027}.
\bibitem{RaSu99}
  {\rm L.~Randall and R.~Sundrum},
  {\it Nucl. Phys.} {\bf B557} {\rm (1999) 79}.
\bibitem{PoRa99}
  {\rm A.~Pomarol and R.~Rattazzi},
  {\it JHEP} {\bf 9905} {\rm (1999) 013}.
\bibitem{GaNeWu99}
  {\rm M.~K.~Gaillard, B.~D.~Nelson and Y.-Y.~Wu},
  {\it Phys. Lett.} {\bf B459} {\rm (1999) 549}.
\bibitem{GaNe00b}
  {\rm M.~K.~Gaillard and B.~D.~Nelson},
  {\it Nucl. Phys.} {\bf B588} {\rm (2000) 197}.
\bibitem{BiGaNe01}
  {\rm P.~Binetruy, M.~K.~Gaillard and B.~D.~Nelson},
  {\it Nucl. Phys.} {\bf B604} {\rm (2001) 32}.
\bibitem{BiGaWu96}
  {\rm P.~Bin\'{e}truy, M.~K.~Gaillard and Y.-Y.~Wu},
  {\it Nucl. Phys.} {\bf B481} {\rm (1996) 109}.
\bibitem{BiGaWu97a}
  {\rm P.~Bin\'{e}truy, M.~K.~Gaillard and Y.-Y.~Wu},
  {\it Nucl. Phys.} {\bf B493} {\rm (1997) 27}.
\bibitem{BrIbMu94}
  {\rm A.~Brignole, C.~E.~Ib\'a\~{n}ez  and  C.~Mu\~{n}oz},
  {\it Nucl. Phys.} {\bf B422} {\rm (1994) 125}.\\
  {\rm ERRATUM}
  {\it Nucl. Phys.} {\bf B436} {\rm (1995) 747}.
\bibitem{DiKaLo91}
  {\rm L.~J.~Dixon, V.~S.~Kaplunovsky and J.~Louis},
  {\it Nucl. Phys.} {\bf B355} {\rm (1991) 649}.
\bibitem{AnNaTa91}
  {\rm I.~Antoniadis, K.~S.~Narain and T.~R.~Taylor},
  {\it Phys. Lett.} {\bf B267} {\rm (1991) 37}.
\bibitem{CaOv93}
  {\rm G.~L.~Cardoso and B.~A.~Ovrut },
  {\it Nucl. Phys.} {\bf B369} {\rm (1993) 351}.
\bibitem{DeFeKoZw92}
  {\rm J.-P.~Derendinger, S.~Ferrara, C.~Kounnas and F.~Zwirner},
  {\it Nucl. Phys.} {\bf B372} {\rm (1992) 145}.
\bibitem{GaTa92}
  {\rm M.~K.~Gaillard and T.~R.~Taylor},
  {\it Nucl. Phys.} {\bf B381} {\rm (1992) 577}.
\bibitem{KaLo95}
  {\rm V.~S.~Kaplunovsky and J.~Louis},
  {\it Nucl. Phys.} {\bf B444} {\rm (1995) 191}.
\bibitem{BiGiGr01}
  {\rm P.~Bin\'{e}truy, G.~Girardi and R.~Grimm},
  {\it Phys. Rept.} {\bf 343} {\rm (2001) 255}.
\bibitem{BiGaWu97b}
  {\rm P.~Bin\'{e}truy, M.~K.~Gaillard and Y.-Y.~Wu},
  {\it Phys. Lett.} {\bf B412} {\rm (1997) 288}.
\bibitem{BaMoPo00}
  {\rm J.~Bagger, T.~Moroi and E.~Poppitz},
  {\it JHEP} {\bf 0004} {\rm (2000) 009}.
\bibitem{snowmass}
  {\rm B.~C.~Allanach et al.},
  {\it Eur. Phys. J.} {\bf C25} {\rm (2001) 113}.
\bibitem{AMSB}
  {\rm R.~Rattazzi, A.~Strumia and J.~D.~Wells},
  {\it Nucl. Phys.} {\bf B576} {\rm (2000) 3}.\\
  {\rm I.~Jack and D.~R.~T.~Jones},
  {\it Phys. Lett.} {\bf B482} {\rm (2000) 167}.\\
%
  {\rm M.~Carena, K.~Huitu and T.~Kobayashi},
  {\it Nucl. Phys.} {\bf B592} {\rm (2001) 164}.
\bibitem{dildom}
  {\rm A.~Brignole, C.~E.~Ib\'a\~{n}ez, C.~Mu\~{n}oz and C.~Scheich},
  {\it Z. Phys.} {\bf C74} {\rm (1997) 157}.\\
  {\rm J.~A.~Casas, A.~Lleyda and C.~Mu\~noz},
  {\it Phys. Lett.} {\bf B380} {\rm (1996) 59}.\\
  {\rm B.~de~Carlos and G.~V.~Kraniotis},
  {\it Relic Abundances and Detection Rates of Neutralinos in String-Inspired
  Supergravity Models},
  {\rm hep-ph/9610355}.\\
  {\rm S.~A.~Abel, B.~C.~Allanach, L.~E.~Ib\'a\~nez, M.~Klein and F.~Quevedo},
  {\it JHEP} {\bf 12} {\rm (2000) 026}.
\bibitem{BaDi94}
  {\rm T.~Banks and M.~Dine},
  {\it Phys. Rev.} {\bf D50} {\rm (1994) 7454}.
\bibitem{Ca96}
  {\rm J.~A.~Casas},
  {\it Phys. Lett.} {\bf B384} {\rm (1996) 103}.
\bibitem{CaLaMuRo90}
  {\rm J.~A.~Casas, Z.~Lalak, C.~Mu\~noz and G.~G.~Ross},
  {\it Nucl. Phys.} {\bf B347} {\rm (1990) 243}.
\bibitem{deCaMu93}
  {\rm B.~de~Carlos, J.~A.~Casas and C.~Mu\~noz},
  {\it Nucl. Phys.} {\bf B399} {\rm (1993) 623}.
\bibitem{Sh90}
  {\rm S.~H.~Shenker},
  {\rm in \it Random Surfaces and Quantum Gravity},
  {\rm Proceedings of the NATO Advanced Study Institute, Cargese, France, 1990},
  {\rm Ed. O.~Alvarez, E.~Marinari and P.~Windey, NATO ASI Series},
  {\rm (Plenum, New York, 1990)}.
\bibitem{BadeCo98}
  {\rm T.~Barreiro, B.~de~Carlos and E.~J.~Copeland},
  {\it Phys. Rev.} {\bf D57} {\rm (1998) 7354}.
\bibitem{Suspect}
  {\rm A.~Djouadi, J.~L.~Kneur and G.~Moultaka},
  {\it SuSpect: a Fortran Code for the Supersymmetric and Higgs Particle
  Spectrum in the MSSM},
  {\rm hep-ph/0211331},
  {\tt http://www.lpm.univ-montp2.fr:6714/\char126kneur/suspect.html}.
\bibitem{GaRiZw90}
  {\rm G.~Gamberini, G.~Ridolfi and F.~Zwirner},
  {\it Nucl. Phys.} {\bf B331} {\rm (1990) 331}.
\bibitem{deCa93}
  {\rm B.~de~Carlos and J.~A.~Casas},
  {\it Phys. Lett.} {\bf B309} {\rm (1993) 320}.
\bibitem{ArNa92}
  {\rm R.~Arnowitt and P.~Nath},
  {\it Phys. Rev.} {\bf D46} {\rm (1992) 3981}.
\bibitem{BaBeOh94}
  {\rm V.~Barger, M.~S.~Berger and P.~Ohmann},
  {\it Phys. Rev.} {\bf D49} {\rm (1994) 4908}.
\bibitem{PiBaMaZh97}
  {\rm D.~Pierce, J.~Bagger, K.~Matchev and R.~Zhang},
  {\it Nucl. Phys.} {\bf B491} {\rm (1997) 3}.
\bibitem{Micromegas}
  {\rm G.~Belanger, F.~Boudjema, A.~Pukhov and A.~Semenov},
  {\it MicrOMEGAs: A Program for Calculating the Relic Density in the MSSM},
  {\rm hep-ph/0112278},\\
  {\tt http://wwwlapp.in2p3.fr/lapth/micromegas}.
\bibitem{LHWG01}
  {\rm LEP Higgs Working Group},
  {\it Searches for the Neutral Higgs Bosons of the MSSM},
  {\rm LHWG Note/2001-04, hep-ex/0107030}.
\bibitem{LHWG02}
  {\rm LEP Higgs Working Group},
  {\it Search for the Standard Model Higgs Boson at LEP},
  {\rm LHWG Note/2002-01}.
\bibitem{Higgslimit}
  {\rm ALEPH Collaboration (A. Heister et al.)}
  {\it Phys.Lett.} {\bf B526} {\rm (2002) 191}.
\bibitem{charginolimit}
  {\rm ALEPH Collaboration (A. Heister et al.)}
  {\it Phys.Lett.} {\bf B533} {\rm (2002) 223}.
\bibitem{Stoplimit}
  {\rm  ALEPH Collaboration (A. Heister et al.)}
  {\it Phys.Lett.} {\bf B537} {\rm (2002) 5}.
\bibitem{KolbTurner} For a review, see: \\
  {\rm E.~W.~Kolb and M.~S.~Turner},
  {\it The Early Universe},
  {\rm Addison-Wesley (New York, 1990)}.
\bibitem{GrSe91}
  {\rm K.~Griest and D.~Seckel},
  {\it Phys. Rev.} {\bf D43} {\rm (1991) 3191};\\
{\rm P.~Binetruy, G.~Girardi and P.~Salati},
{\it Nucl.\ Phys.} {\bf B237} {\rm (1984) 285}.
\bibitem{Stau}
  {\rm J.~R.~Ellis, T.~Falk, K.~A.~Olive and M.~Srednicki},
  {\it Astropart. Phys.} {\bf 13} {\rm (2000) 181}.
\bibitem{Stop}
  {\rm C.~Boehm, A.~Djouadi and M.~Drees},
  {\it Phys. Rev.} {\bf D62} {\rm (2000) 035012};\\
  {\rm J.~R.~Ellis, K.~A.~Olive and Y.~Santoso},
  {\it Astropart. Phys.}  {\bf 18} {\rm (2003) 395}.
\bibitem{Birkedal-Hansen:2001is}
A.~Birkedal-Hansen and B.~D.~Nelson,
{\it Phys. Rev.} {\bf D64} {\rm (2001) 015008}.
\bibitem{Birkedal-Hansen:2002am}
A.~Birkedal-Hansen and B.~D.~Nelson,
{\it Phys. Rev.} {\bf D67} {\rm (2003) 095006}.
\bibitem{EdGo97}
  {\rm J.~Edsjo and P.~Gondolo},
  {\it Phys. Rev.} {\bf D56} {\rm (1997) 1879}.
\bibitem{JuKaGr96}
  {\rm G.~Jungman, M.~Kamionkowski and K.~Griest},
  {\it Phys. Rept.} {\bf 267} {\rm (1996) 195}.
\bibitem{DarkSUSY}
  {\rm P.~Gondolo, J.~Edsjo, L.~Bergstrom, P.~Ullio and E.~A.~Baltz},
  {\it DarkSUSY - A Numerical Package for Dark Matter Calculations in the MSSM},
  {\rm astro-ph/0012234},\\
  {\tt http://www.physto.se/\char126edsjo/darksusy/index.html}.
\bibitem{analytical}
  {\rm T.~Nihei, L.~Roszkowski and R.~R.~de~Austri},
  {\it JHEP} {\bf 0207} {\rm (2002) 024};\\
  {\rm T.~Nihei, L.~Roszkowski and R.~R.~de~Austri},
  {\it JHEP} {\bf 0203} {\rm (2002) 031};\\
  {\rm A.~Birkedal-Hansen and E.~Jeong},
  {\it JHEP} {\bf 0302} {\rm (2003) 047}.
\bibitem{Pr00}
  {\rm J.~R.~Primack},
  {\it Cosmological Parameters 2000},
  {\rm astro-ph/0007187}.
\bibitem{WMAP}
  C.~L.~Bennett {\it et al.}, arXiv:astro-ph/0302207;\\
  D.~N.~Spergel {\it et al.}, arXiv:astro-ph/0302209.

\bibitem{Baer:2003yh}
H.~Baer and C.~Balazs,
arXiv:hep-ph/0303114;
\\
U.~Chattopadhyay, A.~Corsetti and P.~Nath,
arXiv:hep-ph/0303201;
\\
J.~R.~Ellis, K.~A.~Olive, Y.~Santoso and V.~C.~Spanos,
{\it Phys. Lett.} B {\bf B565} {\rm (2003) 176}.

\bibitem{Birkedal-Hansen:2003gy}
A.~Birkedal-Hansen,
arXiv:hep-ph/0306144.

\bibitem{Birkedal-Hansen:2003mp}
A.~Birkedal-Hansen and J.~G.~Wacker,
arXiv:hep-ph/0306161.

\bibitem{Bertolini}
  {\rm S.~Bertolini, F.~Borzumati, A.~Masiero and G.~Ridolfi},
  {\it Nucl. Phys.} {\bf B353} {\rm (1991) 591};\\
  {\rm R.~Barbieri and G.~F.~Giudice},
  {\it Phys. Lett.} {\bf B309} {\rm (1993) 86};\\
  {\rm F.~Borzumati},
  {\it Z. Phys.} {\bf C63} {\rm (1994) 291};\\
  {\rm F.~Borzumati, M.~Olechowski and S.~Pokorski},
  {\it Phys. Lett.} {\bf B349} {\rm (1995) 311};\\
  {\rm G.~Degrassi, P.~Gambino and G.~F.~Giudice},
  {\it JHEP} {\bf 0012} {\rm (2000) 009}.
\bibitem{Cleo}
  {\rm M.~S.~Alam et al. [CLEO Collaboration]},
  {\it Phys. Rev. Lett.} {\bf 74} {\rm (1995) 2885}; \\
  {\rm S.~Ahmed et al.},
  {\rm CLEO CONF 99-10; BELL-CONF-0003},
  {\it Contribution to the 30th International Conference on High-Energy Physics},
  {\rm Osaka, Japan, 2000}.
\bibitem{PDG}
  {\rm Particle Data Group (D.E. Groom et al.)},
  {\it Eur. Phys. J.} {\bf C15} {\rm (2000) 1},
  {\tt http://pdg.lbl.gov/}.
\bibitem{bench}
  {\rm M.~Battaglia, A.~De~Roeck, J.~Ellis, F.~Gianotti, K.~T.~Matchev
  K.~A.~Olive, L.~Pape, G.~Wilson},
  {\it Eur. Phys. J.} {\bf C22} {\rm (2001) 535}.
\bibitem{MaWe03}
  {\rm S.~P.~Martin and J.~D.~Wells},
  {\it Phys. Rev.} {\bf D67} {\rm (2003) 015002}.
\bibitem{Brown}
  {\rm H.~N.~Brown et al. [Muon g-2 Collaboration]},
  {\it Phys. Rev. Lett.} {\bf 86} {\rm (2001) 2227}; \\
  {\rm G.~W.~Bennett et al. [Muon g-2 Collaboration]},
  {\it Phys. Rev. Lett.} {\bf 89} {\rm (2002) 101804}
  {\rm [Erratum-ibid. {\bf 89}, 129903 (2002)]}.
\bibitem{Davier}
  {\rm M.~Davier, S.~Eidelman, A.~Hocker and Z.~Zhang},
  {\rm hep-ph/0208177}.
\bibitem{mSUGRA}
  {\rm J.~R.~Ellis, T.~Falk, G.~Ganis, K.~A.~Olive and M.~Srednicki},
  {\it Phys. Lett.} {\bf B510} {\rm (2001) 236}; \\
  {\rm J.~R.~Ellis, K.~A.~Olive and Y.~Santoso},
  {\it New Jour. Phys.} {\bf 4} {\rm (2002) 32}; \\
  {\rm L.~Roszkowski, R.~Ruiz de Austri and T.~Nihei},
  {\it JHEP} {\bf 0108} {\rm (2001) 024}; \\
  {\rm A.~Djouadi, M.~Drees and J.~L.~Kneur},
  {\it JHEP} {\bf 0108} {\rm (2001) 055}; \\
  {\rm H.~Baer, C.~Balazs and A.~Belyaev},
  {\it JHEP} {\bf 0203} {\rm (2002) 042}.
\bibitem{focus}
  {\rm K.~L.~Chan, U.~Chattopadhyay and P.~Nath},
  {\it Phys. Rev.} {\bf D58} {\rm (1998) 096004}; \\
  {\rm J.~L.~Feng, K.~T.~Matchev and T.~Moroi},
  {\it Phys. Rev. Lett.} {\bf 84} {\rm (2000) 2322}; \\
  {\rm J.~L.~Feng, K.~T.~Matchev and T.~Moroi},
  {\it Phys. Rev.} {\bf D61} {\rm (2000) 075005}.
\bibitem{AlKrPo03}
  {\rm B.~Allanach, S.~Kraml and W.~Porod},
  {\it Theoretical Uncertainties in Sparticle Mass Predictions from
  Computational Tools},
  {\rm hep-ph/0302102}.
\bibitem{AMSBref}
  {\rm T.~Gherghetta, G.~Giudice and J.~D.~Wells},
  {\it Nucl. Phys.} {\bf B559} {\rm (1999) 27}; \\
  {\rm J.~L.~Feng, T.~Moroi, L.~Randall, M.~Strassler and S.~Su},
  {\it Phys. Rev. Lett.} {\bf 83} {\rm (1999) 1731}; \\
  {\rm J.~Feng and T.~Moroi},
  {\it Phys. Rev.} {\bf D61} {\rm (2000) 095004}; \\
  {\rm J.~Gunion and S.~Mrenna},
  {\it Phys. Rev.} {\bf D62} {\rm (2000) 015002}; \\
  {\rm G.~Kribs},
  {\it Phys. Rev.} {\bf D62} {\rm (2000) 015008}; \\
  {\rm H.~Baer, M.~Diaz, P.~Quintana and X.~Tata},
  {\it Phys. Lett.} {\bf B488} {\rm (2000) 367}; \\
  {\rm H.~Baer, J.~K.~Mizukoshi and X.~Tata},
  {\it JHEP} {\bf 0004} {\rm (2000) 016}; \\
  {\rm D.~K.~Ghosh, A.~Kundu, P.~Roy and S.~Roy},
  {\it Phys. Rev.} {\bf D64} {\rm (2001) 115001}; \\
  {\rm J.~Gunion and S.~Mrenna},
  {\it Phys. Rev.} {\bf D64} {\rm (2001) 075002}.
\bibitem{AMSBdm}
  {\rm T.~Moroi and L.~Randall},
  {\it Nucl. Phys.} {\bf B570} {\rm (2000) 455}.
\bibitem{Gi01b}
  {\rm J.~Giedt},
  {\it Annals Phys.} {\bf 289} {\rm (2001) 251}.
\bibitem{KaLyNeWa02}
  {\rm G.~L.~Kane, J.~Lykken, B.~D.~Nelson and L.~Wang},
  {\it Phys. Lett.} {\bf B551} {\rm (2003) 146}.
\bibitem{FoIbLuQu90}
  {\rm A.~Font, L.~Ib\'a\~nez, D.~L\"ust and F.~Quevedo},
  {\it Phys. Lett.} {\bf B245} {\rm (1990) 401}; \\
  {\rm M.~Cvetic, A.~Font, L.~Ib\'a\~nez, D.~L\"ust and F.~Quevedo},
  {\it Nucl. Phys.} {\bf B361} {\rm (1991) 194}.
\bibitem{ourbench}
  {\rm G.~L.~Kane, J.~Lykken, S.~Mrenna, B.~D.~Nelson, L.~Wang and T.~T.~Wang},
  {\it Phys. Rev.} {\bf D67} {\rm (2003) 045008}.
\bibitem{GaNe00a}
  {\rm M.~K.~Gaillard and B.~D.~Nelson},
  {\it Nucl. Phys.} {\bf B571} {\rm (2000) 3}.
\end{thebibliography}
\end{document}